\documentclass[prd,twocolumn,showpacs,preprintnumbers,amsmath,amssymb,superscriptaddress,nofootinbib,english]{revtex4-1}

\usepackage{graphicx}
\usepackage{dcolumn}
\usepackage{bm}
\usepackage{epsfig}
\usepackage{graphicx}
\usepackage{hyperref}
\usepackage[usenames]{color}
\usepackage{url}
\usepackage{enumitem}
\usepackage{float}
\usepackage[utf8]{inputenc}

\hypersetup{
    colorlinks=true,
    linkcolor=blue,
    citecolor=blue,
}

\newcommand{\remove}[1]{}

\def\be{\begin{equation}}
\def\ee{\end{equation}}
\def\ba{\begin{eqnarray}}
\def\ea{\end{eqnarray}}

\def\rd{r_{\rm d}}
\def\rdh{r_{\rm d}h}
\def\om{\Omega_{\rm m}h^2}
\def\Om{\Omega_{\rm m}}

\begin{document}

\title{Recombination--independent determination of the sound horizon and the Hubble constant from BAO}

\author{Levon Pogosian}
\affiliation{Department of Physics, Simon Fraser University, Burnaby, British Columbia, Canada V5A 1S6}
\email[]{levon@sfu.ca}

\author{Gong-Bo Zhao}
\affiliation{National Astronomy Observatories, Chinese Academy of Science, Beijing, 100101, P.R.China}
\affiliation{University of Chinese Academy of Sciences, Beijing, 100049, P.R.China}
\email[]{gbzhao@nao.cas.cn}

\author{Karsten Jedamzik} 
\affiliation{Laboratoire de Univers et Particules de Montpellier, UMR5299-CNRS, Universite de Montpellier, 34095 Montpellier, France}
\email[]{karsten.jedamzik@umontpellier.fr}

\begin{abstract}
The Hubble tension and attempts to resolve it by modifying the physics of (or at) recombination motivate finding ways to determine $H_0$ and the sound horizon at the epoch of baryon decoupling $r_{\rm d}$ in ways that neither rely on a recombination model nor on late-time Hubble data. In this work, we investigate what one can learn from the current and future BAO data when treating $r_{\rm d}$ and $H_0$ as independent free parameters. It is well known that BAO gives exquisite constraints on the product $r_{\rm d}H_0$. We show here that imposing a moderate prior on $\Omega_{\rm m} h^2$ breaks the degeneracy between $r_{\rm d}$ and $H_0$. Using the latest BAO data, including the recently released eBOSS DR16, along with a $\Omega_{\rm m} h^2$ prior based on the Planck best fit $\Lambda$CDM model, we find $r_{\rm d} =143.7 \pm 2.7$ Mpc and $H_0 = 69.6 \pm 1.8$ km/s/Mpc. BAO data therefore prefers somewhat lower $r_{\rm d}$ and higher $H_0$ than those inferred from Planck data in a $\Lambda$CDM model. We find similar values when combing BAO with the Pantheon supernovae, DES galaxy weak lensing, Planck or SPTPol CMB lensing and the cosmic chronometers data. We perform a forecast for DESI and find that, when aided with a moderate prior on $\Omega_{\rm m} h^2$, DESI will measure $r_{\rm d}$ and $H_0$ without assuming a recombination model with an accuracy surpassing the current best estimates from Planck.
\end{abstract}

\maketitle

\section{Introduction}

The 4.2$\sigma$ tension between the Hubble constant $H_0=73.5 \pm 1.4$ km/s/Mpc measured using Supernovae type Ia (SN) calibrated on Cepheid variable stars by the Supernovae H0 for the Equation of State (SH0ES) collaboration \cite{Reid:2019tiq} and the $H_0=67.36 \pm 0.54$ km/s/Mpc value implied by the $\Lambda$ Cold Dark Matter ($\Lambda$CDM) fit to the cosmic microwave background (CMB) anisotropy data from Planck \cite{Aghanim:2018eyx} prompted significant interest in new physics at the epoch of recombination \cite{Poulin:2018cxd,Chiang:2018xpn,Kreisch:2019yzn,Pandey:2019plg,Agrawal:2019lmo,Lin:2019qug,Sakstein:2019fmf,Hart:2019dxi,Jedamzik:2020krr,Gonzalez:2020fdy,Sekiguchi:2020teg} (see \cite{DiValentino:2020zio} for more references). This is because the value of $H_0$ one gets from CMB is directly tied to the sound horizon at last scattering, which is closely related to the sound horizon at the baryon decoupling $\rd$ that sets the characteristic scale of Baryon Acoustic Oscillations  (BAO) in the distribution of large scale structure. Both CMB and BAO measure the angular size of the acoustic scale at the respective redshifts, and a smaller $\rd$ would imply a larger $H_0$. 

While both CMB and BAO determine $H_0$ from the angular acoustic scale, there are some important differences. Firstly, there is much more information in the CMB than just the positions of the acoustic peaks. It is generally not trivial to introduce new physics that reduces $\rd$ without worsening the fit to other features of the  temperature and polarization spectra. Secondly, to get any information about the $H_0$ from CMB spectra, it is not enough to simply know $\rd$ - one actually needs a model of the recombination, since one does not have an independent measure of the redshift of decoupling. In contrast, in the case of the BAO, one knows the redshift of the BAO feature from spectroscopy of galaxies, so there is more hope of learning something about the $H_0$ without relying on a recombination model. 

It is well-known that BAO observations constrain the product $\rdh$, where $h \equiv H_0$/(100 km/s/Mpc)\footnote{We refer only to the measurements of the BAO peaks, not the full shape of the galaxy power spectrum. The latter also carries the imprint of the scale of the horizon at the radiation-matter equality \cite{Philcox:2020xbv}.}. Several strategies have been adopted to break the degeneracy between $\rd$ and $h$, while avoiding using information from CMB spectra (which is based on a recombination model and measures both $\rd$ and $h$ exquisitely well). One option is to assume a particular recombination model, supplemented by a prior on the baryon density \cite{Addison:2013haa,Addison:2017fdm,Wang:2017yfu,Cuceu:2019for,DAmico:2019fhj,Ivanov:2019pdj,Philcox:2020vvt,Alam:2020sor}, which is well-constrained by the Big Bang Nucleosynthesis (BBN) \cite{Cyburt:2015mya}. This places a prior on $\rd$ which then helps to constrain $H_0$. Further combining BAO and BBN with weak lensing (WL) and SN data results in tight constraints on cosmological parameters \cite{Abbott:2017smn}. Alternatively, one can combine BAO with measurements of the Hubble constant  to infer $\rd$~\cite{Aylor:2018drw,Wojtak:2019tkz,Arendse:2019hev}. Neither strategy is fully satisfactory as it is either model-dependent or relies on observational data which is in tension. In fact the latter method simply recasts the Hubble tension as the $\rd$ tension. Since solutions to the Hubble tension include proposals of modified recombination, it would be preferable to have a recombination-model-independent determination of both $\rd$ and $H_0$ using datasets that are not in tension with either SH0ES or Planck. We show that this is indeed possible.

As we show in Sec.~\ref{sec2}, a prior on $\om$ helps to break the degeneracy between $\rd$ and $H_0$. Hence, treating $r_d$ and $H_0$ as independent observables, in Sec.~\ref{sec:data} we combine BAO with data capable of constraining $\om$, such as galaxy and CMB WL. The CMB lensing power spectra are particularly useful as they probe the largest scales of the underlying matter power spectrum, including the horizon scale at the matter-radiation equality \cite{Baxter:2020qlr}. Following \cite{Zhang:2020uan}, we also include the cosmic chronometer (OHD) data \cite{Magana:2017nfs}. In addition, we derive bounds on $\rd$ and $H_0$ from BAO alone supplemented by a moderate prior on $\om$. Moderate means that it is sufficiently weak to be consistent with the Planck best fit model as well as viable models with modified recombination histories. 

Interestingly, we find that both methods, using BAO+data and BAO+prior, give almost identical mean values for $\rd$ and $H_0$ and similar $1\sigma$ uncertainties around $3$ Mpc and $1.7$-$1.8$ km/s/Mpc, respectively. We find the mean value of $H_0$ to be around $69.5$ km/s/Mpc, lying in between the Planck and the SH0ES values. Thus, we find that, when no recombination model is assumed, the BAO data is not in significant tension with either of the two.

Furthermore we perform a forecast for future BAO data from the Dark Energy Spectroscopic Instrument (DESI) and show that, when combined with a moderate prior on $\om$ it will constrain $\rd$ and $H_0$ with precision better than Planck's, without the need for a recombination model. Future CMB experiments, such as the Simons Observatory (SO) \cite{Ade:2018sbj} and CMB-S4 \cite{Abazajian:2016yjj}, will significantly improve on the current CMB lensing reconstructions \cite{Mirmelstein:2019sxi} and can be used along with the future galaxy WL data from Euclid \cite{euclid} and Legacy Survey of Space and Time (LSST) \cite{lsst}. Thus, we expect excellent recombination independent bounds on $\rd$ and $H_0$ from the combination of DESI, SO/CMB-S4 and Euclid/LSST, but leave the detailed forecast to a future study.

Finally we emphasize the importance of the $\rdh$ parameter, which can be well-measured by BAO alone along with $\Om$. As we show, current BAO data measures $\rdh$ to a percent level accuracy. It agrees well with the $\Lambda$CDM value derived from Planck and is in tension with some alternative models. DESI will measure $\rdh$ and $\Om$ with accuracy 4-5 times better than Planck's in a recombination independent way, providing a powerful consistency test capable of falsifying competing models.

\section{BAO observables and the parameter degeneracies}
\label{sec2}

\begin{figure*}[htbp] 
\includegraphics[scale=0.45]{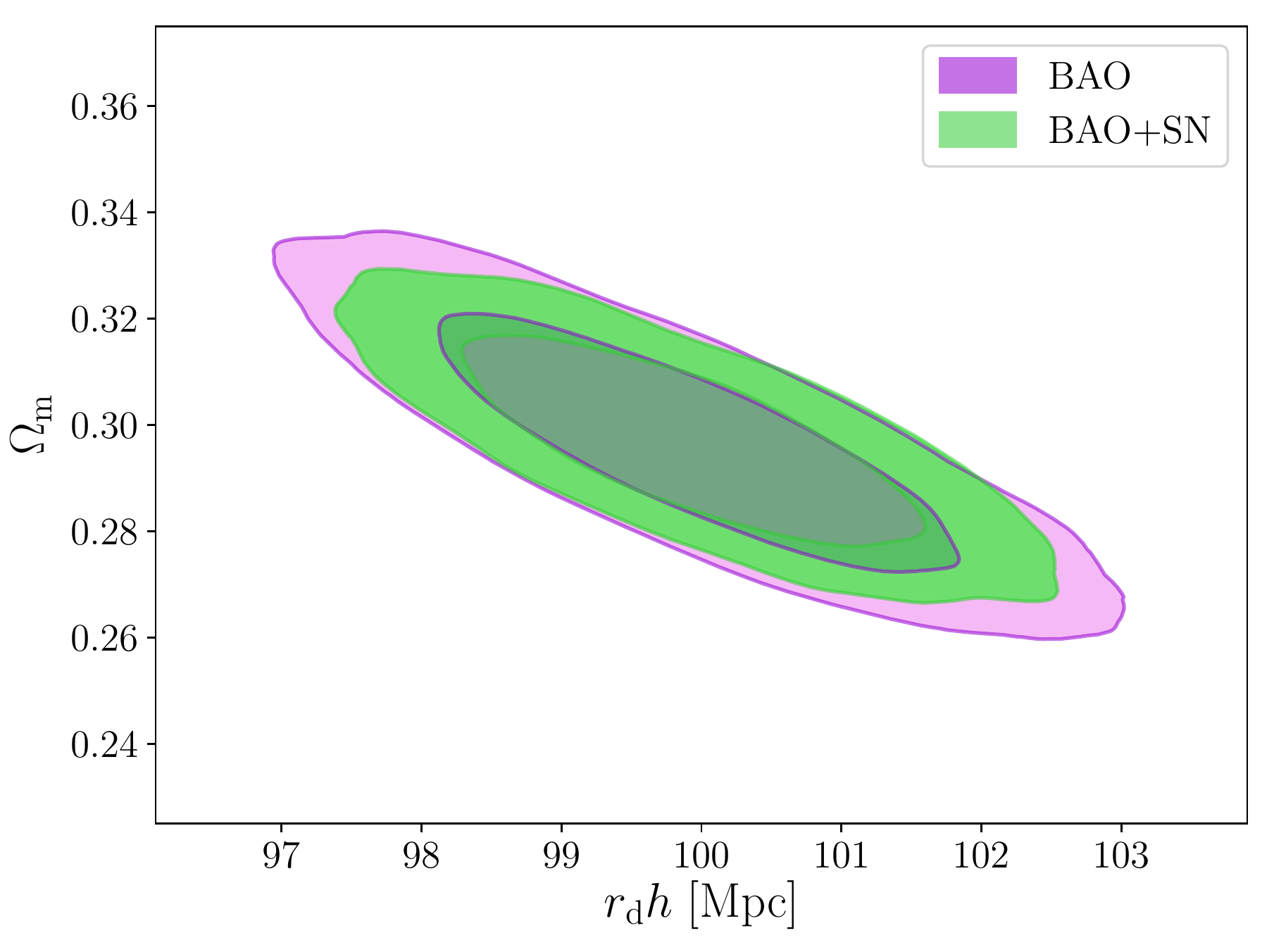}
\includegraphics[scale=0.45]{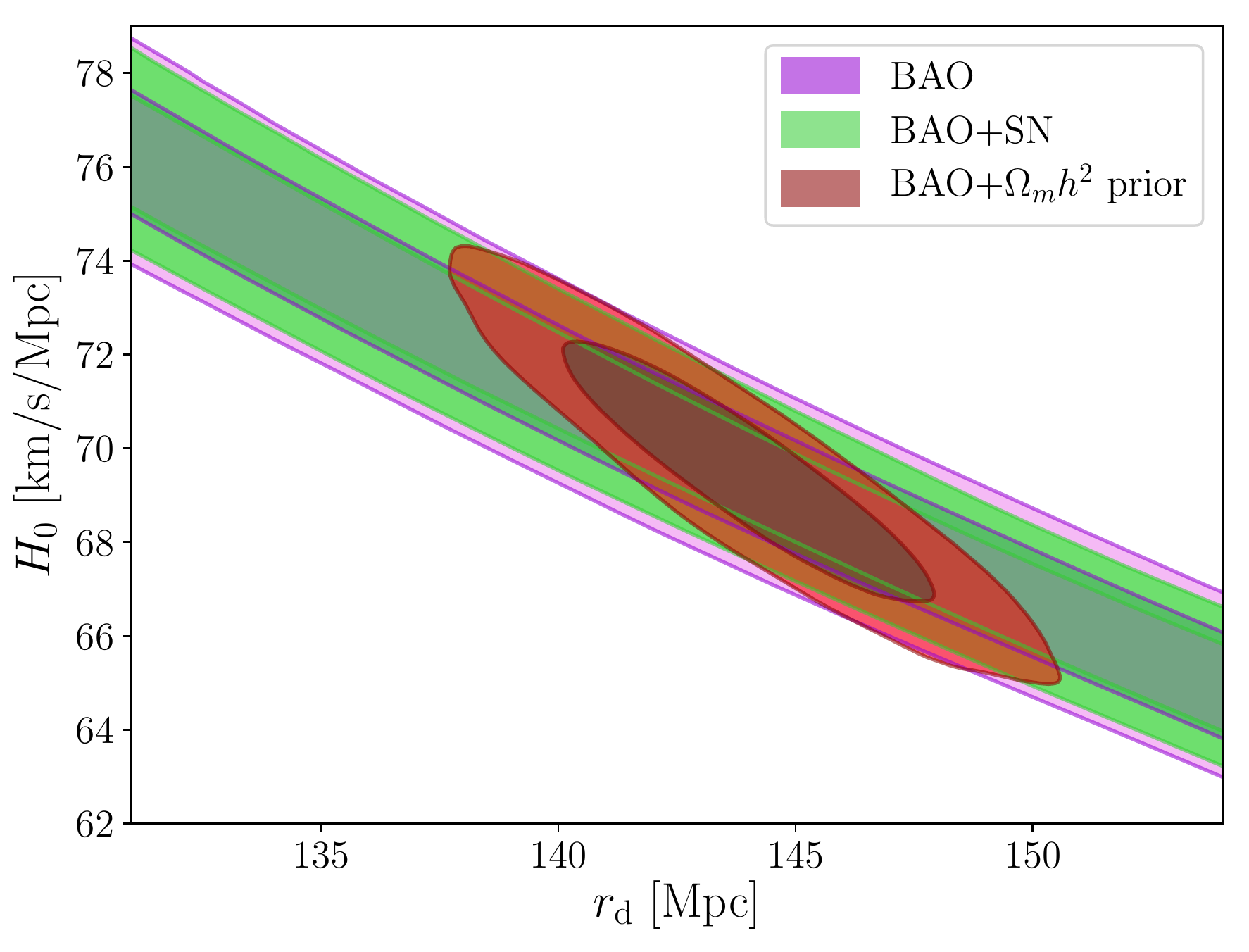}
\caption{Constraints on $\rdh$, $\Om$, $\rd$ and $H_0$ derived from the latest combination of the BAO data, and from BAO combined with SN. The right panel shows that a prior on $\om$ breaks the degeneracy between $\rd$ and $H_0$.} 
\label{fig:bao}
\end{figure*} 

The BAO scale is set by the comoving sound horizon $\rd$ at the epoch of baryon decoupling $z_{\rm d}$, also called the ``drag'' epoch\footnote{It is closely related to the sound horizon at last scattering, $r_\star  \approx 1.02 \rd$ \cite{Anderson:2013zyy,Aubourg:2014yra}, that sets the positions of the acoustic peaks in the CMB spectra.}. It is given by
\be
\rd = \int^\infty_{z_{\rm d}} {c_S(z) {\rm d} z \over H(z)} ,
\label{eq:rd}
\ee
where the sound speed $c_S(z)$ is a known function of the baryon to photon densities ratio, and
\be
H(z) = H_0 \sqrt{\Omega_r(1+z)^4+\Om(1+z)^3+1-\Om-\Omega_r} ,
\label{eq:bigH}
\ee
where $\Omega_r$ and $\Om$ denote the present day density fractions in relativistic and non-relativistic matter, and where we have assumed a flat $\Lambda$CDM universe, so that $\Omega_\Lambda = 1-\Om-\Omega_r$. It is often useful to work in terms of the dimensionless Hubble parameter $h(z) = H(z)/$(100 km/s/Mpc) and the physical density parameters $\omega_i \equiv \Omega_i h^2$, where $h \equiv h(0)$. Rewriting Eq.~(\ref{eq:bigH}) as
\be
h(z) = 
\sqrt{\omega_r (1+z)^4+\omega_m(1+z)^3 + h^2 - \omega_m-\omega_r}
\label{eq:h}
\ee
makes it apparent that, once the physical densities $\omega_i$ are provided, the value of $h$ plays practically no role at $z > z_{\rm d} \sim 1000$ and, hence, in the integral (\ref{eq:rd}). This justifies treating $\rd$ as a parameter independent of $H_0$.

The BAO observables one extracts from surveys of galaxies and other tracers of large scale structure are of three types  \cite{Eisenstein:2005su}: the acoustic feature measured using correlations in the direction perpendicular to the line of sight,
\be
\beta_\perp(z) = D_M(z)/\rd ,
\ee
where $D_M(z) = \int_0^z c{\rm d}z'/H(z')$ is the comoving distance to redshift $z$, the feature measured in the direction parallel to the line of sight,
\be
\beta_\parallel(z) = H(z) \rd ,
\ee
and the angle-averaged or ``isotropic'' measurement,
\be
\beta_V(z) = D_V(z)/\rd , 
\ee
where $D_V(z) = \left[ czD_M^2(z)/H(z) \right]^{1/3}$. At redshifts of relevance to galaxy surveys we can safely ignore the contribution of relativistic species in the expression for $h(z)$. Then, $\beta_\perp$ can be written as
\be
\beta_\perp(z) = \int_0^z {2998 \ {\rm Mpc} \ {\rm d}z' \over \rdh \sqrt{\Om(1+z')^3 + 1- \Om}} 
\label{beta_rdh}
\ee
or, equivalently, as
\be
\beta_\perp(z)
=\int_0^z {2998 \ {\rm Mpc} \ {\rm d} z'  \over \rd \omega_m^{1/2}\sqrt{(1+z')^3 + h^2/\omega_m - 1}} .
\label{hrd_omh2}
\ee
From Eq.~(\ref{beta_rdh}) it is clear that having BAO measurements at multiple redshifts allows one to measure two numbers: $\rdh$ and $\Om$. It is also evident from Eq.~(\ref{hrd_omh2}) that one can break the degeneracy between $\rd$ and $h$ by supplementing BAO with a prior on $\omega_m$. The same argument also applies to the other two BAO observables. 

Fig.~\ref{fig:bao} illustrates the above points. In the left panel we show the constraints on $\rdh$ and $\Om$ derived from the latest BAO data (detailed in Sec.~\ref{sec:data}), while the right panel shows the corresponding bounds on $\rd$ and $H_0$. Adding the SN data helps to constrain $\Om$, thus slightly reducing the uncertainties in the $\rdh-\Om$ plane. As one can see from the right panel, adding a prior on $\om$ breaks the degeneracy allowing to constrain $\rd$ and $H_0$ individually.

\section{Constraints from current data}
\label{sec:data}

\begin{figure}[tbp] 
\includegraphics[scale=0.45]{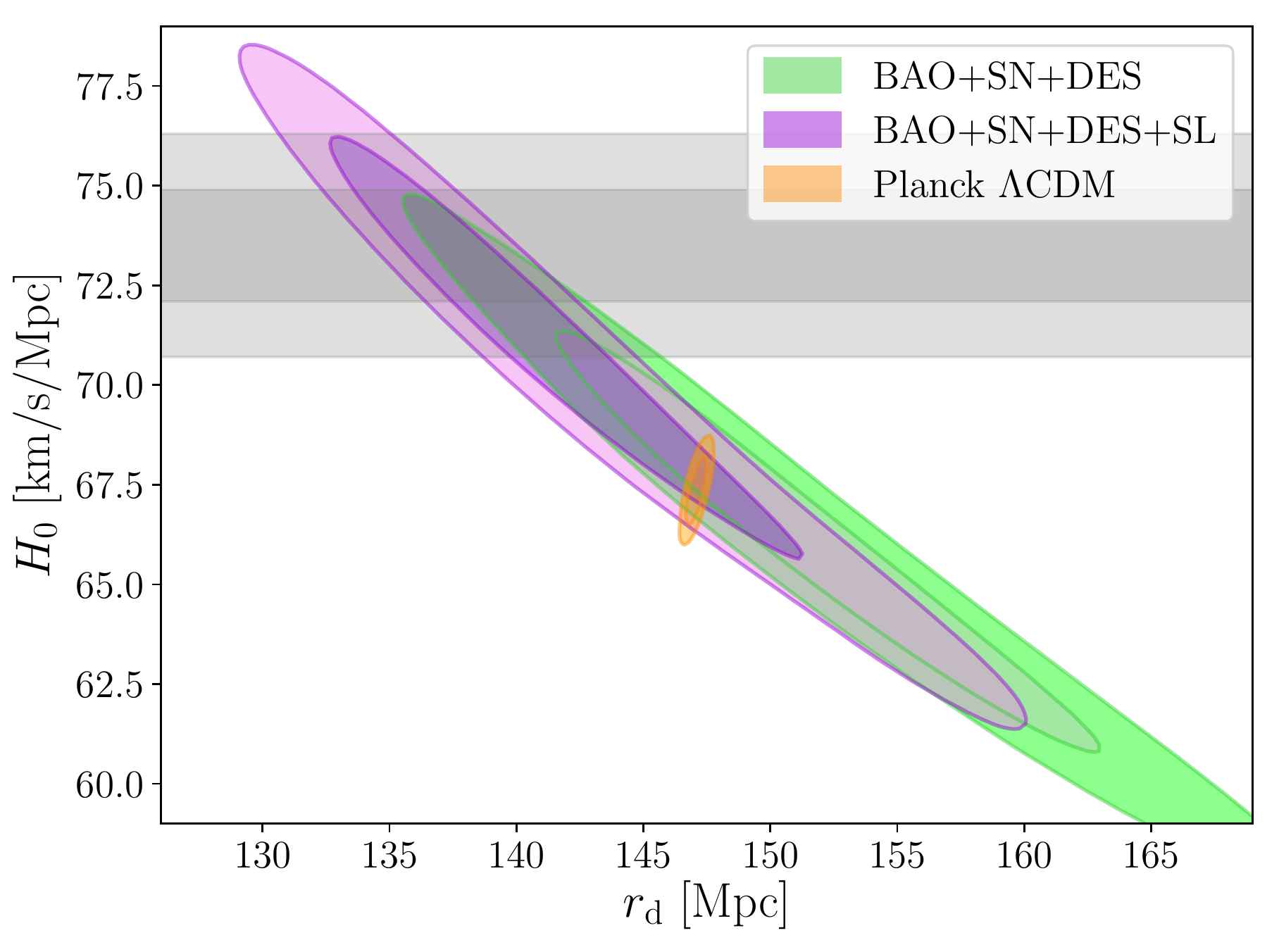}
\includegraphics[scale=0.45]{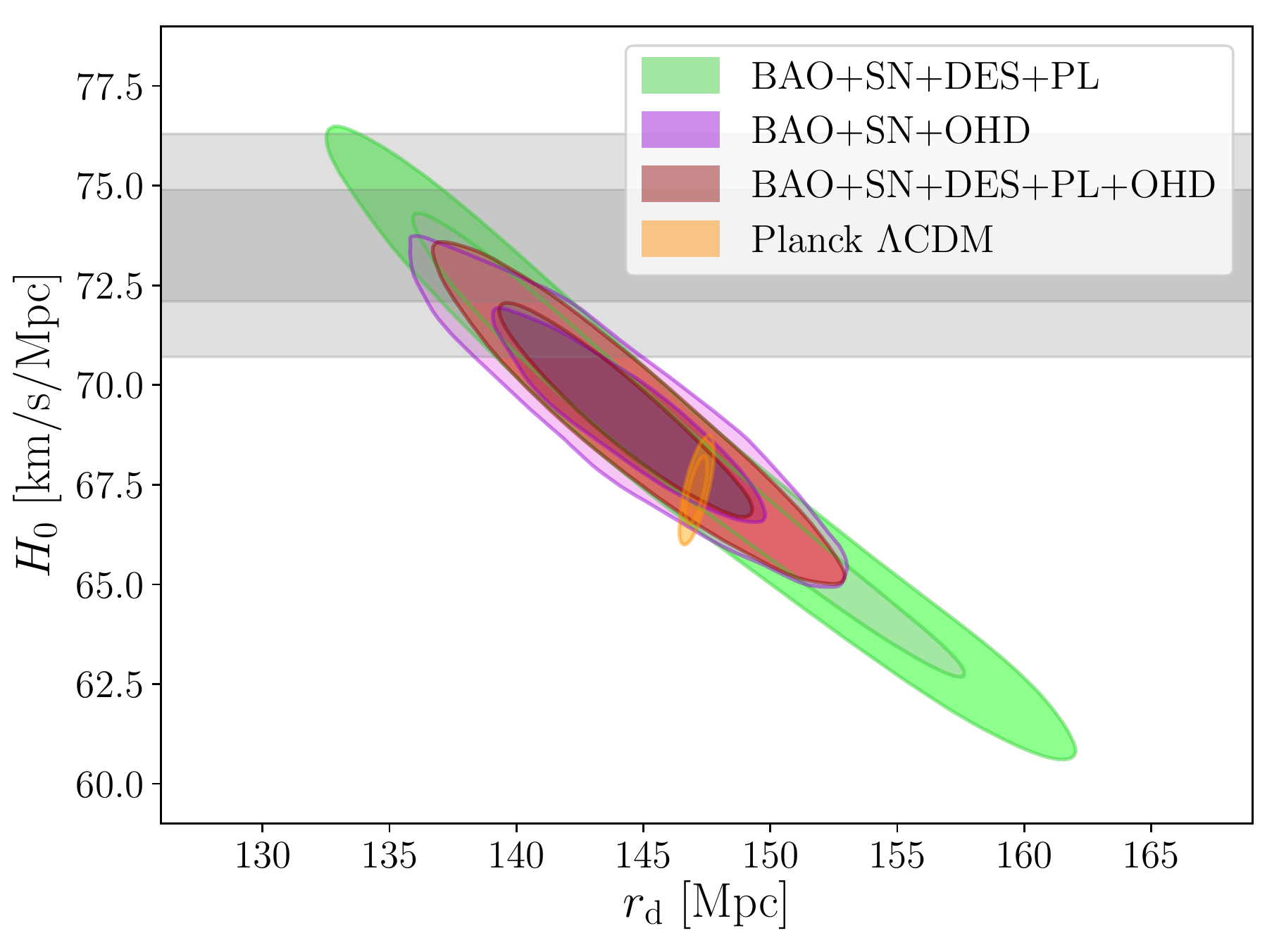}
\caption{Constraints on $\rd$ and $H_0$ derived from the BAO data combined with Pantheon SN, DES galaxy WL, CMB WL from Planck and SPTPol, and the OHD data. The grey bands show the 68\% and 95\% CL determination of $H_0$ by SH0ES. The $\Lambda$CDM based bound from Planck CMB anisotropy spectra is shown for reference.} \label{fig:bshwl}
\end{figure} 

\begin{figure}[htbp]
\includegraphics[scale=0.45]{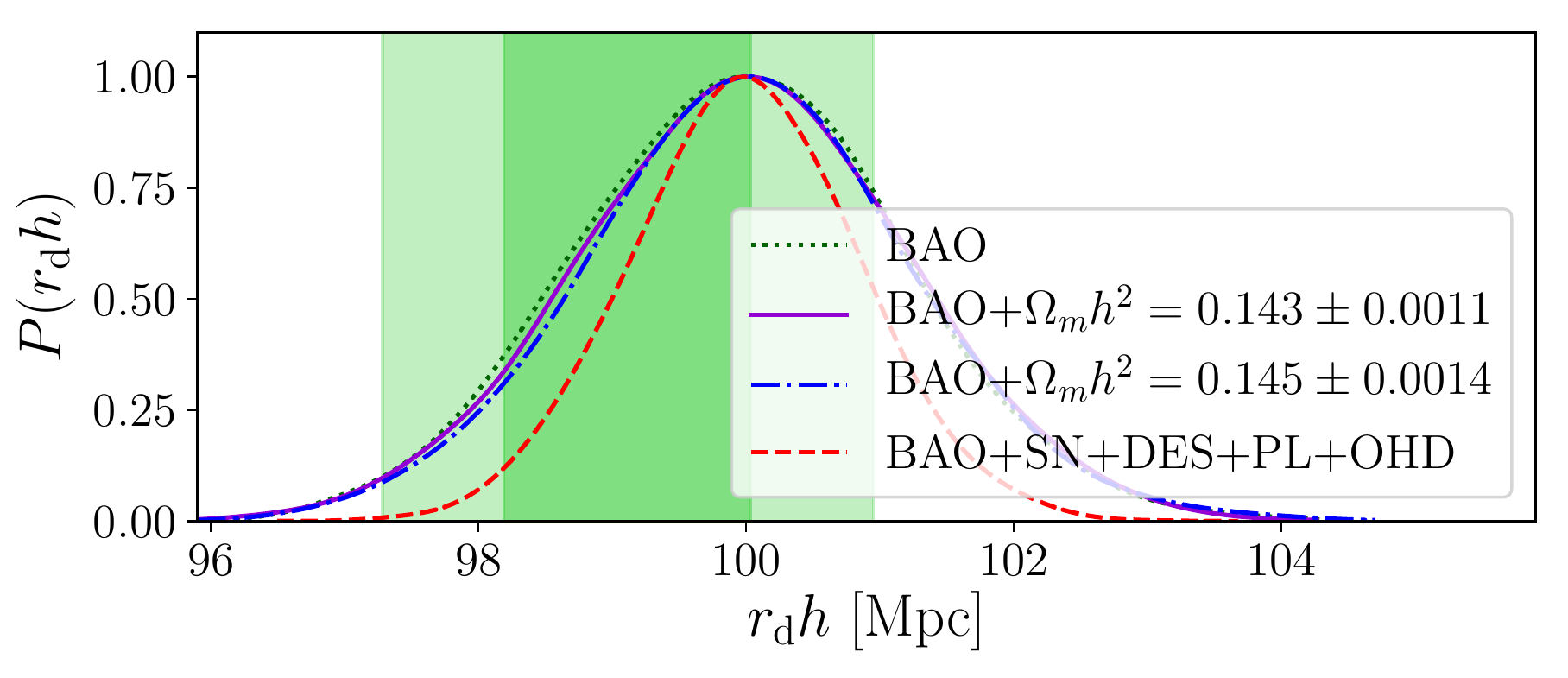}
\includegraphics[scale=0.45]{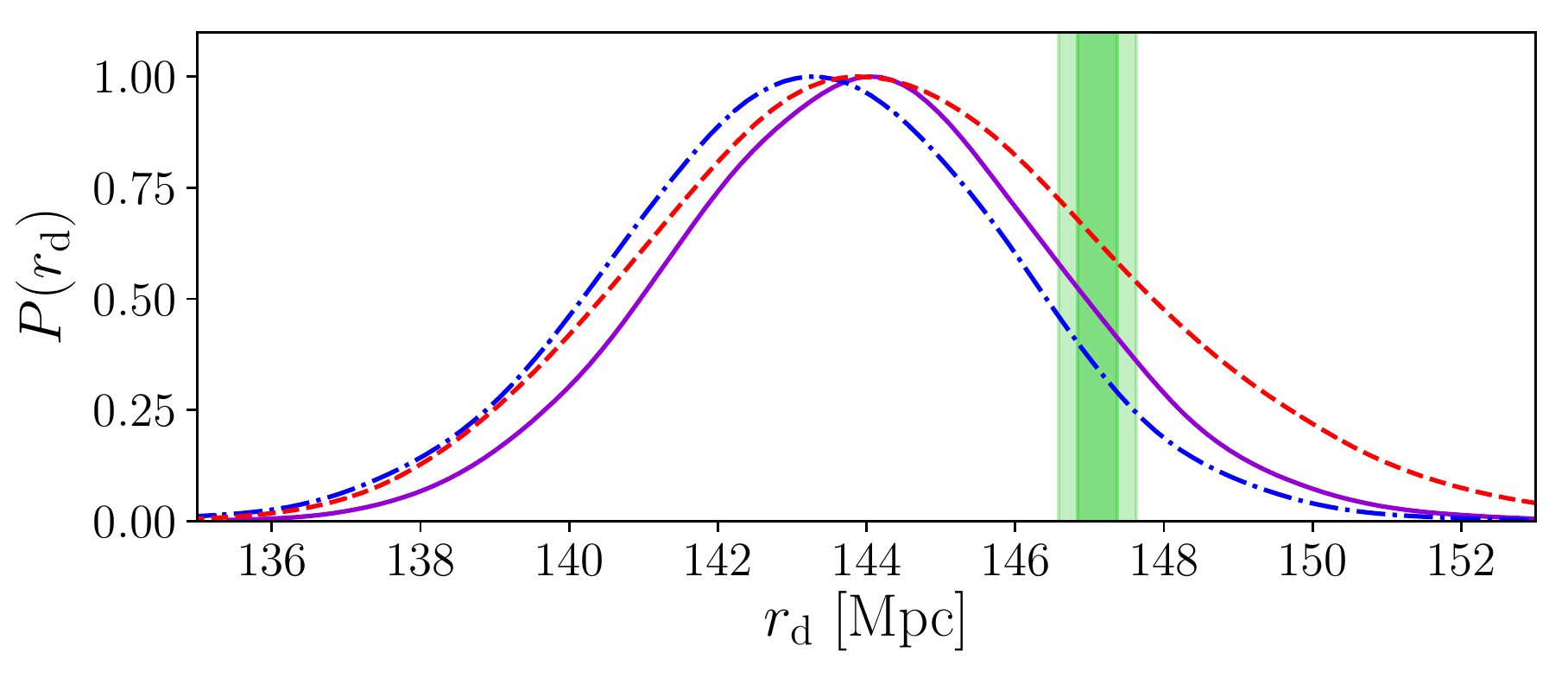}
\includegraphics[scale=0.45]{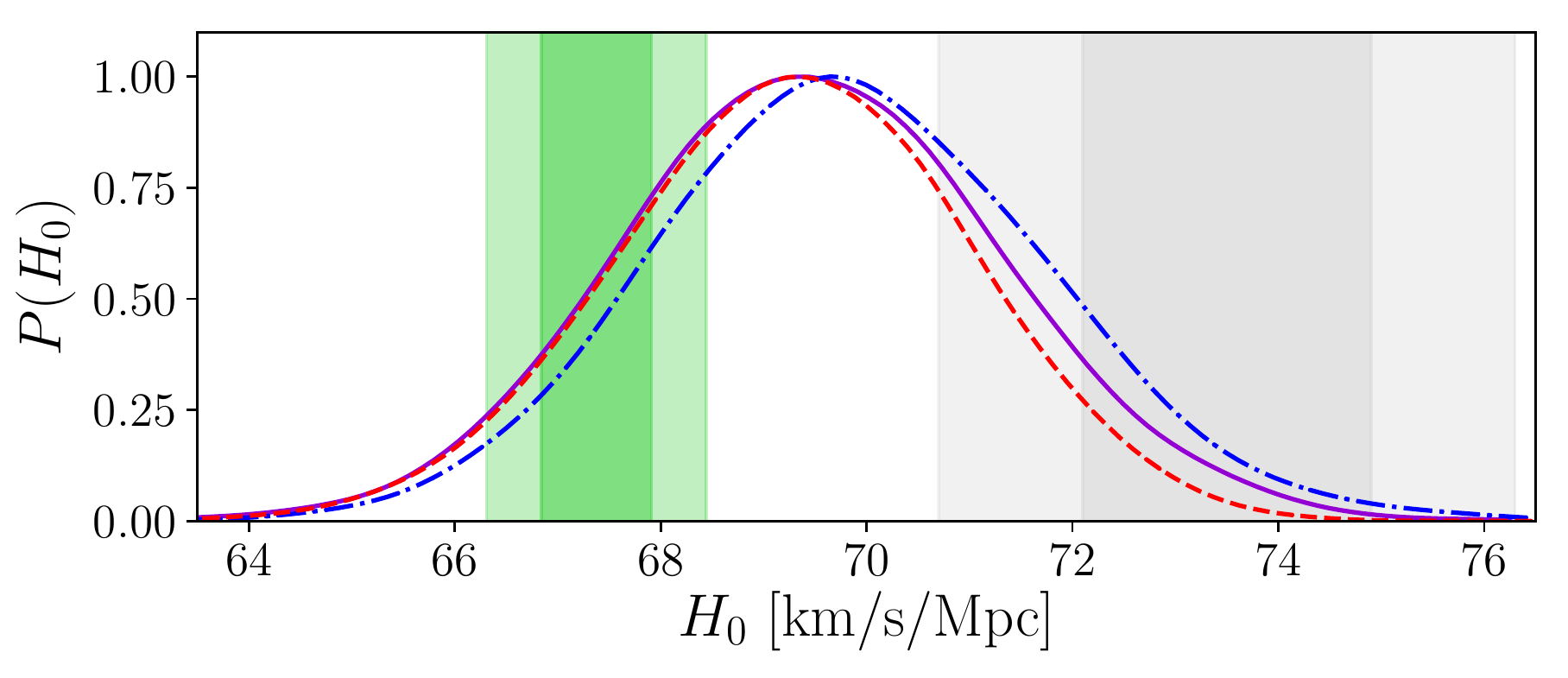}
\caption{Constraints on $r_{\rm d}h$, $\rd$ and $H_0$ from BAO and different priors on $\om$. The constraint from a combination of current recombination-independent data is shown as well. The green vertical bands correspond to the Planck best fit $\Lambda$CDM. The grey bands show the $H_0$ measurement by SH0ES. }\label{fig:prior1d}
\end{figure} 

\begin{table*}[htbp]
\centering
\begin{tabular}{c|c|c|c|c}
\hline\hline
        & $r_{\rm d} h$ [Mpc] & $\Om$  & $\rd$ [Mpc] &  $H_0$ [km/s/Mpc] \\
\hline
BAO & $99.95\pm 1.2$ & $0.297^{+0.014}_{-0.016}$ & - &  -  \\
BAO+SN & $99.9\pm 1.0$ & $0.297\pm 0.013$ & - &  -  \\
BAO+SN+DES & $100.1\pm 1.0$ & $0.294^{+0.011}_{-0.012}$ & $152.3^{+6.2}_{-7.6}$ &  $65.9\pm 3.4$  \\
BAO+SN+PL & $100.0\pm 1.0$ & $0.295\pm 0.012$ & $151.3^{+7.8}_{-4.2}$ &  $66.2^{+1.8}_{-3.6}$  \\
BAO+SN+SL & $99.9\pm 1.1$ & $0.297\pm 0.013$ & $145.5^{+6.7}_{-9.1}$ &  $68.8\pm 3.6$  \\
BAO+SN+OHD & $99.9\pm 1.0$ & $0.298\pm 0.013$ & $144.4\pm 3.4$ &  $69.2\pm 1.7$  \\
BAO+SN+DES+PL & $99.9\pm 1.0$ & $0.297\pm 0.012$ & $145.9^{+5.0}_{-8.3}$ &  $68.6^{+4.1}_{-3.2}$  \\
BAO+SN+DES+SL & $100.2\pm 1.0$ & $0.292^{+0.011}_{-0.014}$ & $142.1^{+4.0}_{-7.3} $ &  $70.7^{+4.0}_{-2.7}$  \\
BAO+SN+DES+PL+OHD & $99.99\pm 0.84$ & $0.2961\pm 0.0083$ & $144.4^{+2.8}_{-3.4}$ &   $69.3\pm 1.7$  \\
BAO+SN+DES+SL+OHD & $99.96\pm 0.85$ & $0.2960\pm 0.0083$ & $143.6^{+2.8}_{-3.3}$ &   $69.6\pm 1.7$  \\
BAO+fixed $\om=0.143$ & $100.1\pm 1.2$  & $0.294^{+0.014}_{-0.016}$ & $143.7\pm 2.5$   &  $69.7\pm 1.8$  \\
BAO+prior $\om=0.143 \pm 0.0011$ & $100.0\pm 1.2$   & $0.294^{+0.014}_{-0.016}$  & $143.8 \pm 2.6$  &  $69.6\pm 1.9 $  \\
BAO+prior $\om=0.143 \pm 0.0022$ & $99.99\pm 1.2$   & $0.294^{+0.014}_{-0.016}$  & $143.7\pm 2.7$  &  $69.6\pm 1.8$  \\
BAO+fixed $\om=0.145$ & $99.95\pm 1.2$   & $0.295\pm 0.016$  & $142.9\pm 2.5$  &  $70.0\pm 1.8$  \\
BAO+prior $\om=0.145 \pm 0.0014$ & $100.0\pm 1.2$   & $0.294^{+0.015}_{-0.017}$  & $142.9\pm 2.6$  &  $70.0\pm 1.9$  \\
BAO+prior $\om=0.145 \pm 0.0028$ & $100.0\pm 1.2$  & $0.294^{+0.014}_{-0.016} $  & $142.7\pm 2.8$  &  $70.1\pm 1.9$  \\
\hline\hline
\end{tabular}
\caption{\label{tab:params} The mean parameter values and 68\% CL uncertainties derived from the considered combinations of datasets.}
\end{table*}

We use a collection of BAO measurements to date, including the ones derived from the recently released Date Release (DR) 16 of the extended Baryon Oscillation Spectroscopic Survey (eBOSS) \cite{Alam:2020sor}. Being a multi-tracer galaxy survey, eBOSS provides BAO and redshift space distortions (RSD) measurements at multiple redshifts from the samples of Luminous Red Galaxies (LRGs), Emission Line Galaxies (ELGs), clustering quasars (QSOs), and the Lyman-$\alpha$ forest. In this work, we use the BAO measurement from the full-shape auto- and cross-power spectrum of the eBOSS LRGs and ELGs \cite{Zhao:2020tis,Wang:2020tje}, the BAO measurement from the QSO sample \cite{Hou:2020rse}, and from the Lyman-$\alpha$ forest sample \cite{duMasdesBourboux:2020pck}. Since all these measurements are at $z>0.6$, we combine with low-$z$ measurements, including the BAO measurement by 6dF \cite{Beutler:2011hx}, SDSS DR7 main Galaxy sample (MGS) \cite{Ross:2014qpa}, to complement. 

As explained in Sec.~\ref{sec2}, BAO on their own can constrain $\Om$ and the product $\rdh$. To constrain $\rd$ and $H_0$ individually, one can either supplement BAO with data that provides a prior on $\om$, or data that constrains $H_0$, or the combination of the two. Restricting to datasets that do not rely on modelling the recombination physics, the first option includes the galaxy and the CMB weak lensing data. To that aim, we consider the Dark Energy Survey Year 1 galaxy clustering and weak lensing data (DES) \cite{Abbott:2017wau}, and the CMB lensing power spectra from Planck 2018 (PL) \cite{Aghanim:2018oex} and SPTpol (SL) \cite{Wu:2019hek,Bianchini:2019vxp}. Both types of measurements are practically insensitive to the scale of baryon decoupling and primarily probe the cumulative clustering of matter. While, in principle, a different redshift of decoupling would change the time at which baryons begin to cluster, this is a very minor effect on the net growth of cosmic structures dominated by dark matter. For the second option, to avoid data contributing to the Hubble tension, we use the cosmic chronometer data (OHD) from \cite{OHD:Moresco_2016,Ratsimbazafy:2017vga}. The latter contain determinations of $H(z)$ at 31 redshifts in the $0.1\lesssim z \lesssim 2$ range and, since $\Om$ and $H_0$ are the only parameters in our flat FRW model, provides a handle on the value of $H_0$ when combined with the BAO.

We use {\tt CosmoMC} \cite{Lewis:2002ah} modified to work with $\rd$ as an independent parameter. The cosmological parameters we vary are $\rd$, $H_0$ and either $\Om$ or $\om$. When using the DES and CMB lensing data, we additionally vary the amplitude of the primordial fluctuations spectrum $A_s$ and the spectral index $n_s$. As was shown in \cite{Ade:2015zua}, CMB lensing constrains the combination of $A_s(\Om^{0.6}h)^{2.3}$, where $A_s$ is the primordial fluctuations spectrum amplitude. Further combining it with galaxy lensing helps to constrain $A_s$ and deliver a prior on $\om$. We also use the Pantheon SN sample \cite{Scolnic:2017caz} which does not help in breaking the $\rd$-$H_0$ degeneracy but still helps a little bit by providing an independent constraint on $\Om$. We find that the combination of the SN, DES and PL data gives $\om  = 0.140\pm 0.011$ at 68\% confidence level (CL). This constraint is an order of magnitude weaker than that derived from the Planck CMB anisotropies, but future weak lensing data will do significantly better. 

Fig.~\ref{fig:bshwl} shows the effect of combing BAO with weak lensing data, namely BAO+SN+DES, BAO+SN+DES+PL and BAO+SN+DES+SL, and with the OHD data, as well as their combination. The comprehensive list of parameter constraints from various data combinations is given in Table~\ref{tab:params}. Clearly, the OHD data dominates the constraints when included. We also note that BAO+SN+SL prefers a somewhat higher $H_0$ and smaller $\rd$, while still being quite consistent with BAO+SN+PL. The mean values obtained from BAO+SN+OHD and BAO+SN+DES+PL(+SL) also show a good consistency with each other, although the uncertainties in the latter are large. Combining all the data together, we find $H_0=69.3/69.6 \pm 1.7$ and $\rd = 144.4/143.6^{+2.8}_{-3.4/3.3}$ from BAO+SN+DES+PL/SL+OHD.

In addition to analyzing the above-mentioned combinations of datasets, we separately consider the BAO data supplemented by several externally imposed Gaussian priors on $\om$. Fig.~\ref{fig:prior1d} shows the posterior distributions of the relevant parameters derived using two choices of priors: one based on the Planck best fit $\Lambda$CDM \cite{Aghanim:2018eyx} and the other on an alternative recombination model that also gives an acceptable fit to the CMB \cite{Jedamzik:2020krr}. The plot shows that the results are not very sensitive to the choice of the prior. Table~\ref{tab:params} shows results with two different fixed values of $\om$, and priors of doubled width, all giving comparable outcomes. This indicates that the uncertainties are dominated by those in the current BAO data. As we will see in the next section, the strength of the $\om$ prior will play a more important role for future BAO data from DESI.

Imposing a prior on $\om$ is, to some extent, a matter of choice. While $\om$ has the well-defined physical meaning of the present day matter density, imposing a prior on another combinations of $\Om$ and $h$ would also do the job.  In fact, a study dedicated to the consistency test between CMB and BAO could benefit from combining the latter with a prior on $\Om h^3$, which is the combination best constrained by CMB in a flat FRW cosmology \cite{Percival:2002gq}. We leave exploring this possibility to a separate investigation.

It is worth noting that the CMB-derived best fit value of $\om$ is quite consistent between a number of models with modified recombination histories\footnote{It is notably larger in early dark energy (EDE) models, which puts them in tension with the galaxy weak lensing data \cite{Hill:2020osr,Ivanov:2020ril,DAmico:2020ods,Ye:2020oix} (see also \cite{Murgia:2020ryi,Smith:2020rxx} for an alternative perspective).}. This further justifies applying a prior on $\om$ when attempting to gain recombination-model-independent information from BAO. It also provides a consistency test with the results obtained by combining BAO with the weak lensing and the OHD data.

For comparison, in Fig.~\ref{fig:prior1d}, we also show the constraints from the BAO+SN+DES+PL/SL+OHD which are largely the same as those derived using the $\om$ prior. Both methods give $r_{\rm d} \approx 144 \pm 3$ Mpc and $H_0 \approx 69.5 \pm 1.8$ km/s/Mpc. The latter number is in between and within the 1-2 $\sigma$ range of the Planck and SH0ES values, shown with vertical bands.

As the top panel in Fig.~\ref{fig:prior1d} shows, BAO alone can constrain the product $r_{\rm d}h$ to a percent level accuracy, yielding a value that is in a perfect agreement with Planck's $\Lambda$CDM. We note that $\rdh$, if measured with sufficient accuracy, can be used to discriminate between models. As we will show in the forecast section, DESI alone will determine $\rdh$ and $\Om$ with accuracy several times better than Planck's providing a powerful consistency test without any assumptions about recombination physics.

\section{Forecast for DESI}
\label{sec:forecast}

\begin{figure*}[tbp] 
\includegraphics[scale=0.3]{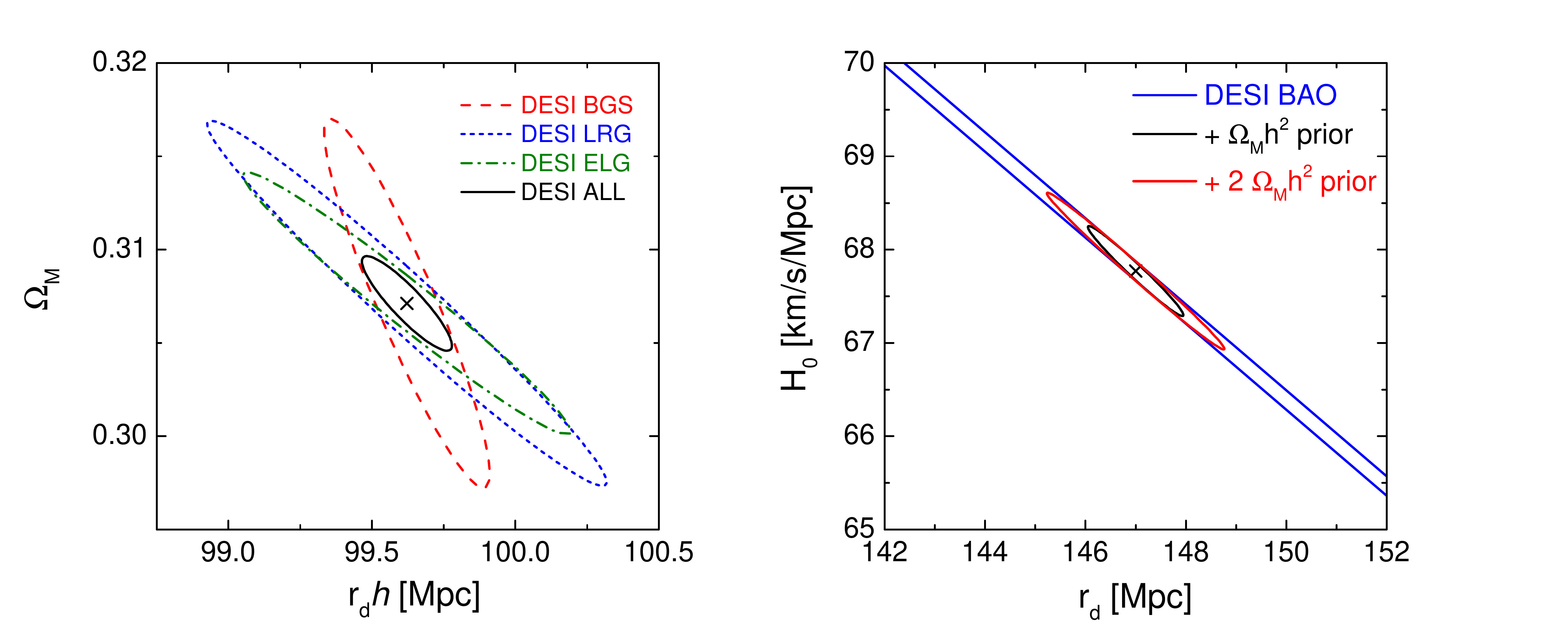}
\caption{Forecasted constraints on $\{\rdh,\Om\}$ (left) and $\{\rd,H_0\}$ (right) using specification of the DESI survey. The contours represent the 68\% CL constraint, and the crosses mark the fiducial model. In the right panel, different priors on $\om$ are applied, where `$\om$ \ prior' means the Gaussian prior on $\om$ from the Planck 2018 observations, namely, $\sigma_{\om}=0.0011$, and `$2\om$ \ prior' means the $2\sigma$ Planck prior.} \label{fig:forecast}
\end{figure*} 

In this section, we perform a Fisher forecast for $\rd$ and $H_0$, as well as $\rdh$ and $\Om$, using specifications of the Dark Energy Spectroscopic Instrument (DESI) \citep{DESI}, an upcoming stage-IV galaxy survey. We assume that DESI will survey $14,000$ deg$^2$ of the sky, using the Bright Galaxies (BGS) ($z\lesssim0.45$), Luminous Red Galaxies (LRGs) ($0.65 \lesssim z \lesssim1.15$) and Emission Line Galaxies (ELGs) ($0.65\lesssim z \lesssim1.65$), and that we are able to extract the tomographic information on the past light cone at a redshift resolution of $\Delta z \lesssim 0.1$. In this forecast, we used the full-shape anisotropic galaxy power spectrum as observable, and marginalize over the RSD, bias and Fingers-of-God parameters.

\begin{table}[htbp]
\begin{center}
\begin{tabular}{c|cccccc}
\hline\hline
Parameter & BGS & LRG & ELG & ALL & +$\sigma(\omega_m)$ & +$2\sigma(\omega_m)$\\
\hline
$\sigma(\rdh)$ & $0.192$  & $0.464$ & $0.380$ & $0.105$ & - & - \\
$\sigma(\Om)$ &  $0.0066$ & $0.0065$ & $0.0047$ & $0.0017$  & - & - \\
\hline
$\sigma(\rd)$ & - & - & - & - & $0.636$ & $1.179$ \\
$\sigma(H_0)$ & - & - & - & - & $0.323$ &  $0.560$ \\
\hline\hline
\end{tabular}
\end{center}
\label{tab:forecast}
\caption{A forecast for $\{\rdh,\Om\}$ and $\{\rd,H_0\}$ using different DESI tracers with Planck priors on $\om$.}
\end{table}%

The forecast results are shown in Fig.~\ref{fig:forecast} and Table~\ref{tab:forecast}. We find that DESI alone will be able to measure $\rdh$ in a recombination-independent way with an accuracy of $\sim 0.1\%$, almost an order of magnitude better than Planck, providing an important consistency check. DESI will also constrain $\Om$ with a five-fold improvement in accuracy over Planck.

With the help of a Gaussian prior on $\om$, based on the present estimate from Planck \cite{Aghanim:2018eyx},
DESI will measure $\rd$ and $H_0$ with $1\sigma$ uncertainties of $\sim 0.6$ Mpc and $\sim 0.3$ km/s/Mpc, respectively. Unlike the case with the current BAO data, which we saw not to be particularly sensitive to the width of the $\om$ priors, the results from DESI will be directly dependent on it. As Fig.~\ref{fig:forecast} and Table~\ref{tab:forecast} show, doubling the width of the prior doubles the uncertainties in $\rd$ and $H_0$. Even then, DESI would yield results with accuracy comparable to Planck's.

The sensitivity of DESI to the prior on $\om$ prompts one to seek alternatives ways to constrain it to a similar accuracy. The additional information could come from the CMB weak lensing spectra from SO and CMB-S4, which will improve considerably on Planck lensing \cite{Mirmelstein:2019sxi}, as well as galaxy lensing from Euclid and LSST.

\section{Summary}

We have shown that there is a wealth of information that one can extract from the BAO data without using information that depends on a particular recombination model.  In particular, one can measure $\rd$ and $H_0$ from the BAO by supplementing it with a prior on $\om$. This can be done by combining BAO with the lensing information from either the CMB or the galaxies, or imposing a moderate Gaussian prior based on a consensus determination of $\om$ from CMB. We find that the combination of BAO, SN, OHD, DES and PL (or SL) give competitive determinations of both parameters, with $r_{\rm d} \approx 144.4^{+2.8}_{-3.4}$ ($143.6^{+2.8}_{-3.3}$ ) Mpc and $H_0 \approx 69.3 \pm 1.7 (69.6 \pm 1.7)$ km/s/Mpc, showing an excellent consistency with $r_{\rm d} \approx 143.8 \pm 2.6$ Mpc and $H_0 \approx 69.6\pm 1.9$ km/s/Mpc obtained using the BAO+prior method. They are also consistent at $1\sigma$ level with the Planck best fit $\Lambda$CDM values of $\rd = 147.10 \pm 0.27$ Mpc and $H_0 = 67.37 \pm 0.54$ km/s/Mpc.

We found that current BAO data provides a competitive constraint on the product $\rdh$, showing a good agreement with the best fit $\Lambda$CDM value from Planck. We have also performed a forecast for DESI, finding that it will constrain $\rdh$ and $\Om$ with an order of magnitude better accuracy that will allow for a powerful consistency check against parameters determined from CMB.

Future CMB experiments, like the SO \cite{Ade:2018sbj} and CMB-S4 \cite{Abazajian:2016yjj} will significantly improve on the current CMB lensing reconstructions \cite{Mirmelstein:2019sxi}, while Euclid and LSST will provide much better galaxy lensing data. It will be interesting to perform a detailed forecast for DESI+SO/S4+Euclid/LSST using $\rd$ as an independent variable. We leave this to a future study.

It is evident that a recombination-model-independent determination of $r_{\rm d}$ and $H_0$ prefers somewhat larger $H_0$ and smaller $r_{\rm d}$ than Planck data under the assumption of $\Lambda$CDM. Such values of $H_0$ are also consistent with the $H_0$ determination from the tip of the red giant branch \cite{Freedman:2019jwv}. However, a smaller tension with SH0ES still remains. There seems to be enough theory space for modifications of the cosmological recombination process which is consistent with these inferred values of $r_{\rm d}$ and $H_0$ \cite{DiValentino:2020zio}. Future data will show if indeed it is necessary to amend $\Lambda$CDM. 

\acknowledgments

We thank Nikki Arendse and Eiichiro Komatsu for useful discussions, and the anonymous referee for comments that helped to improve the paper. We gratefully acknowledge using {\tt GetDist} \cite{Lewis:2019xzd}. This research was enabled in part by support provided by WestGrid ({\tt www.westgrid.ca}) and Compute Canada Calcul Canada ({\tt www.computecanada.ca}). L.P. is supported in part by the National Sciences and Engineering Research Council (NSERC) of Canada, and by the Chinese Academy of Sciences President's International Fellowship Initiative, Grant No. 2020VMA0020. G.B.Z. is supported by the National Key Basic Research and Development Program of China (No. 2018YFA0404503), a grant of CAS Interdisciplinary Innovation Team, and NSFC Grants 11925303, 11720101004, 11673025 and 11890691.


\begin{thebibliography}{64}%
\makeatletter
\providecommand \@ifxundefined [1]{%
 \@ifx{#1\undefined}
}%
\providecommand \@ifnum [1]{%
 \ifnum #1\expandafter \@firstoftwo
 \else \expandafter \@secondoftwo
 \fi
}%
\providecommand \@ifx [1]{%
 \ifx #1\expandafter \@firstoftwo
 \else \expandafter \@secondoftwo
 \fi
}%
\providecommand \natexlab [1]{#1}%
\providecommand \enquote  [1]{``#1''}%
\providecommand \bibnamefont  [1]{#1}%
\providecommand \bibfnamefont [1]{#1}%
\providecommand \citenamefont [1]{#1}%
\providecommand \href@noop [0]{\@secondoftwo}%
\providecommand \href [0]{\begingroup \@sanitize@url \@href}%
\providecommand \@href[1]{\@@startlink{#1}\@@href}%
\providecommand \@@href[1]{\endgroup#1\@@endlink}%
\providecommand \@sanitize@url [0]{\catcode `\\12\catcode `\$12\catcode
  `\&12\catcode `\#12\catcode `\^12\catcode `\_12\catcode `\%12\relax}%
\providecommand \@@startlink[1]{}%
\providecommand \@@endlink[0]{}%
\providecommand \url  [0]{\begingroup\@sanitize@url \@url }%
\providecommand \@url [1]{\endgroup\@href {#1}{\urlprefix }}%
\providecommand \urlprefix  [0]{URL }%
\providecommand \Eprint [0]{\href }%
\providecommand \doibase [0]{http://dx.doi.org/}%
\providecommand \selectlanguage [0]{\@gobble}%
\providecommand \bibinfo  [0]{\@secondoftwo}%
\providecommand \bibfield  [0]{\@secondoftwo}%
\providecommand \translation [1]{[#1]}%
\providecommand \BibitemOpen [0]{}%
\providecommand \bibitemStop [0]{}%
\providecommand \bibitemNoStop [0]{.\EOS\space}%
\providecommand \EOS [0]{\spacefactor3000\relax}%
\providecommand \BibitemShut  [1]{\csname bibitem#1\endcsname}%
\let\auto@bib@innerbib\@empty
\bibitem [{\citenamefont {Reid}\ \emph {et~al.}(2019)\citenamefont {Reid},
  \citenamefont {Pesce},\ and\ \citenamefont {Riess}}]{Reid:2019tiq}%
  \BibitemOpen
  \bibfield  {author} {\bibinfo {author} {\bibfnamefont {M.}~\bibnamefont
  {Reid}}, \bibinfo {author} {\bibfnamefont {D.}~\bibnamefont {Pesce}}, \ and\
  \bibinfo {author} {\bibfnamefont {A.}~\bibnamefont {Riess}},\ }\href
  {\doibase 10.3847/2041-8213/ab552d} {\bibfield  {journal} {\bibinfo
  {journal} {Astrophys. J. Lett.}\ }\textbf {\bibinfo {volume} {886}},\
  \bibinfo {pages} {L27} (\bibinfo {year} {2019})},\ \Eprint
  {http://arxiv.org/abs/1908.05625} {arXiv:1908.05625 [astro-ph.GA]}
  \BibitemShut {NoStop}%
\bibitem [{\citenamefont {Aghanim}\ \emph
  {et~al.}(2018{\natexlab{a}})\citenamefont {Aghanim} \emph
  {et~al.}}]{Aghanim:2018eyx}%
  \BibitemOpen
  \bibfield  {author} {\bibinfo {author} {\bibfnamefont {N.}~\bibnamefont
  {Aghanim}} \emph {et~al.} (\bibinfo {collaboration} {Planck}),\ }\href@noop
  {} {\  (\bibinfo {year} {2018}{\natexlab{a}})},\ \Eprint
  {http://arxiv.org/abs/1807.06209} {arXiv:1807.06209 [astro-ph.CO]}
  \BibitemShut {NoStop}%
\bibitem [{\citenamefont {Poulin}\ \emph {et~al.}(2019)\citenamefont {Poulin},
  \citenamefont {Smith}, \citenamefont {Karwal},\ and\ \citenamefont
  {Kamionkowski}}]{Poulin:2018cxd}%
  \BibitemOpen
  \bibfield  {author} {\bibinfo {author} {\bibfnamefont {V.}~\bibnamefont
  {Poulin}}, \bibinfo {author} {\bibfnamefont {T.~L.}\ \bibnamefont {Smith}},
  \bibinfo {author} {\bibfnamefont {T.}~\bibnamefont {Karwal}}, \ and\ \bibinfo
  {author} {\bibfnamefont {M.}~\bibnamefont {Kamionkowski}},\ }\href {\doibase
  10.1103/PhysRevLett.122.221301} {\bibfield  {journal} {\bibinfo  {journal}
  {Phys. Rev. Lett.}\ }\textbf {\bibinfo {volume} {122}},\ \bibinfo {pages}
  {221301} (\bibinfo {year} {2019})},\ \Eprint
  {http://arxiv.org/abs/1811.04083} {arXiv:1811.04083 [astro-ph.CO]}
  \BibitemShut {NoStop}%
\bibitem [{\citenamefont {Chiang}\ and\ \citenamefont
  {Slosar}(2018)}]{Chiang:2018xpn}%
  \BibitemOpen
  \bibfield  {author} {\bibinfo {author} {\bibfnamefont {C.-T.}\ \bibnamefont
  {Chiang}}\ and\ \bibinfo {author} {\bibfnamefont {A.}~\bibnamefont
  {Slosar}},\ }\href@noop {} {\  (\bibinfo {year} {2018})},\ \Eprint
  {http://arxiv.org/abs/1811.03624} {arXiv:1811.03624 [astro-ph.CO]}
  \BibitemShut {NoStop}%
\bibitem [{\citenamefont {Kreisch}\ \emph {et~al.}(2019)\citenamefont
  {Kreisch}, \citenamefont {Cyr-Racine},\ and\ \citenamefont
  {Dore}}]{Kreisch:2019yzn}%
  \BibitemOpen
  \bibfield  {author} {\bibinfo {author} {\bibfnamefont {C.~D.}\ \bibnamefont
  {Kreisch}}, \bibinfo {author} {\bibfnamefont {F.-Y.}\ \bibnamefont
  {Cyr-Racine}}, \ and\ \bibinfo {author} {\bibfnamefont {O.}~\bibnamefont
  {Dore}},\ }\href@noop {} {\  (\bibinfo {year} {2019})},\ \Eprint
  {http://arxiv.org/abs/1902.00534} {arXiv:1902.00534 [astro-ph.CO]}
  \BibitemShut {NoStop}%
\bibitem [{\citenamefont {Pandey}\ \emph {et~al.}(2020)\citenamefont {Pandey},
  \citenamefont {Karwal},\ and\ \citenamefont {Das}}]{Pandey:2019plg}%
  \BibitemOpen
  \bibfield  {author} {\bibinfo {author} {\bibfnamefont {K.~L.}\ \bibnamefont
  {Pandey}}, \bibinfo {author} {\bibfnamefont {T.}~\bibnamefont {Karwal}}, \
  and\ \bibinfo {author} {\bibfnamefont {S.}~\bibnamefont {Das}},\ }\href
  {\doibase 10.1088/1475-7516/2020/07/026} {\bibfield  {journal} {\bibinfo
  {journal} {JCAP}\ }\textbf {\bibinfo {volume} {07}},\ \bibinfo {pages} {026}
  (\bibinfo {year} {2020})},\ \Eprint {http://arxiv.org/abs/1902.10636}
  {arXiv:1902.10636 [astro-ph.CO]} \BibitemShut {NoStop}%
\bibitem [{\citenamefont {Agrawal}\ \emph {et~al.}(2019)\citenamefont
  {Agrawal}, \citenamefont {Cyr-Racine}, \citenamefont {Pinner},\ and\
  \citenamefont {Randall}}]{Agrawal:2019lmo}%
  \BibitemOpen
  \bibfield  {author} {\bibinfo {author} {\bibfnamefont {P.}~\bibnamefont
  {Agrawal}}, \bibinfo {author} {\bibfnamefont {F.-Y.}\ \bibnamefont
  {Cyr-Racine}}, \bibinfo {author} {\bibfnamefont {D.}~\bibnamefont {Pinner}},
  \ and\ \bibinfo {author} {\bibfnamefont {L.}~\bibnamefont {Randall}},\
  }\href@noop {} {\  (\bibinfo {year} {2019})},\ \Eprint
  {http://arxiv.org/abs/1904.01016} {arXiv:1904.01016 [astro-ph.CO]}
  \BibitemShut {NoStop}%
\bibitem [{\citenamefont {Lin}\ \emph {et~al.}(2019)\citenamefont {Lin},
  \citenamefont {Benevento}, \citenamefont {Hu},\ and\ \citenamefont
  {Raveri}}]{Lin:2019qug}%
  \BibitemOpen
  \bibfield  {author} {\bibinfo {author} {\bibfnamefont {M.-X.}\ \bibnamefont
  {Lin}}, \bibinfo {author} {\bibfnamefont {G.}~\bibnamefont {Benevento}},
  \bibinfo {author} {\bibfnamefont {W.}~\bibnamefont {Hu}}, \ and\ \bibinfo
  {author} {\bibfnamefont {M.}~\bibnamefont {Raveri}},\ }\href@noop {} {\
  (\bibinfo {year} {2019})},\ \Eprint {http://arxiv.org/abs/1905.12618}
  {arXiv:1905.12618 [astro-ph.CO]} \BibitemShut {NoStop}%
\bibitem [{\citenamefont {Sakstein}\ and\ \citenamefont
  {Trodden}(2019)}]{Sakstein:2019fmf}%
  \BibitemOpen
  \bibfield  {author} {\bibinfo {author} {\bibfnamefont {J.}~\bibnamefont
  {Sakstein}}\ and\ \bibinfo {author} {\bibfnamefont {M.}~\bibnamefont
  {Trodden}},\ }\href@noop {} {\  (\bibinfo {year} {2019})},\ \Eprint
  {http://arxiv.org/abs/1911.11760} {arXiv:1911.11760 [astro-ph.CO]}
  \BibitemShut {NoStop}%
\bibitem [{\citenamefont {Hart}\ and\ \citenamefont
  {Chluba}(2020)}]{Hart:2019dxi}%
  \BibitemOpen
  \bibfield  {author} {\bibinfo {author} {\bibfnamefont {L.}~\bibnamefont
  {Hart}}\ and\ \bibinfo {author} {\bibfnamefont {J.}~\bibnamefont {Chluba}},\
  }\href {\doibase 10.1093/mnras/staa412} {\bibfield  {journal} {\bibinfo
  {journal} {Mon. Not. Roy. Astron. Soc.}\ }\textbf {\bibinfo {volume} {493}},\
  \bibinfo {pages} {3255} (\bibinfo {year} {2020})},\ \Eprint
  {http://arxiv.org/abs/1912.03986} {arXiv:1912.03986 [astro-ph.CO]}
  \BibitemShut {NoStop}%
\bibitem [{\citenamefont {Jedamzik}\ and\ \citenamefont
  {Pogosian}(2020)}]{Jedamzik:2020krr}%
  \BibitemOpen
  \bibfield  {author} {\bibinfo {author} {\bibfnamefont {K.}~\bibnamefont
  {Jedamzik}}\ and\ \bibinfo {author} {\bibfnamefont {L.}~\bibnamefont
  {Pogosian}},\ }\href@noop {} {\  (\bibinfo {year} {2020})},\ \Eprint
  {http://arxiv.org/abs/2004.09487} {arXiv:2004.09487 [astro-ph.CO]}
  \BibitemShut {NoStop}%
\bibitem [{\citenamefont {Gonzalez}\ \emph {et~al.}(2020)\citenamefont
  {Gonzalez}, \citenamefont {Hertzberg},\ and\ \citenamefont
  {Rompineve}}]{Gonzalez:2020fdy}%
  \BibitemOpen
  \bibfield  {author} {\bibinfo {author} {\bibfnamefont {M.}~\bibnamefont
  {Gonzalez}}, \bibinfo {author} {\bibfnamefont {M.~P.}\ \bibnamefont
  {Hertzberg}}, \ and\ \bibinfo {author} {\bibfnamefont {F.}~\bibnamefont
  {Rompineve}},\ }\href@noop {} {\  (\bibinfo {year} {2020})},\ \Eprint
  {http://arxiv.org/abs/2006.13959} {arXiv:2006.13959 [astro-ph.CO]}
  \BibitemShut {NoStop}%
\bibitem [{\citenamefont {Sekiguchi}\ and\ \citenamefont
  {Takahashi}(2020)}]{Sekiguchi:2020teg}%
  \BibitemOpen
  \bibfield  {author} {\bibinfo {author} {\bibfnamefont {T.}~\bibnamefont
  {Sekiguchi}}\ and\ \bibinfo {author} {\bibfnamefont {T.}~\bibnamefont
  {Takahashi}},\ }\href@noop {} {\  (\bibinfo {year} {2020})},\ \Eprint
  {http://arxiv.org/abs/2007.03381} {arXiv:2007.03381 [astro-ph.CO]}
  \BibitemShut {NoStop}%
\bibitem [{\citenamefont {Di~Valentino}\ \emph {et~al.}(2020)\citenamefont
  {Di~Valentino} \emph {et~al.}}]{DiValentino:2020zio}%
  \BibitemOpen
  \bibfield  {author} {\bibinfo {author} {\bibfnamefont {E.}~\bibnamefont
  {Di~Valentino}} \emph {et~al.},\ }\href@noop {} {\  (\bibinfo {year}
  {2020})},\ \Eprint {http://arxiv.org/abs/2008.11284} {arXiv:2008.11284
  [astro-ph.CO]} \BibitemShut {NoStop}%
\bibitem [{\citenamefont {Philcox}\ \emph
  {et~al.}(2020{\natexlab{a}})\citenamefont {Philcox}, \citenamefont {Sherwin},
  \citenamefont {Farren},\ and\ \citenamefont {Baxter}}]{Philcox:2020xbv}%
  \BibitemOpen
  \bibfield  {author} {\bibinfo {author} {\bibfnamefont {O.~H.}\ \bibnamefont
  {Philcox}}, \bibinfo {author} {\bibfnamefont {B.~D.}\ \bibnamefont
  {Sherwin}}, \bibinfo {author} {\bibfnamefont {G.~S.}\ \bibnamefont {Farren}},
  \ and\ \bibinfo {author} {\bibfnamefont {E.~J.}\ \bibnamefont {Baxter}},\
  }\href@noop {} {\  (\bibinfo {year} {2020}{\natexlab{a}})},\ \Eprint
  {http://arxiv.org/abs/2008.08084} {arXiv:2008.08084 [astro-ph.CO]}
  \BibitemShut {NoStop}%
\bibitem [{\citenamefont {Addison}\ \emph {et~al.}(2013)\citenamefont
  {Addison}, \citenamefont {Hinshaw},\ and\ \citenamefont
  {Halpern}}]{Addison:2013haa}%
  \BibitemOpen
  \bibfield  {author} {\bibinfo {author} {\bibfnamefont {G.~E.}\ \bibnamefont
  {Addison}}, \bibinfo {author} {\bibfnamefont {G.}~\bibnamefont {Hinshaw}}, \
  and\ \bibinfo {author} {\bibfnamefont {M.}~\bibnamefont {Halpern}},\ }\href
  {\doibase 10.1093/mnras/stt1687} {\bibfield  {journal} {\bibinfo  {journal}
  {Mon. Not. Roy. Astron. Soc.}\ }\textbf {\bibinfo {volume} {436}},\ \bibinfo
  {pages} {1674} (\bibinfo {year} {2013})},\ \Eprint
  {http://arxiv.org/abs/1304.6984} {arXiv:1304.6984 [astro-ph.CO]} \BibitemShut
  {NoStop}%
\bibitem [{\citenamefont {Addison}\ \emph {et~al.}(2018)\citenamefont
  {Addison}, \citenamefont {Watts}, \citenamefont {Bennett}, \citenamefont
  {Halpern}, \citenamefont {Hinshaw},\ and\ \citenamefont
  {Weiland}}]{Addison:2017fdm}%
  \BibitemOpen
  \bibfield  {author} {\bibinfo {author} {\bibfnamefont {G.}~\bibnamefont
  {Addison}}, \bibinfo {author} {\bibfnamefont {D.}~\bibnamefont {Watts}},
  \bibinfo {author} {\bibfnamefont {C.}~\bibnamefont {Bennett}}, \bibinfo
  {author} {\bibfnamefont {M.}~\bibnamefont {Halpern}}, \bibinfo {author}
  {\bibfnamefont {G.}~\bibnamefont {Hinshaw}}, \ and\ \bibinfo {author}
  {\bibfnamefont {J.}~\bibnamefont {Weiland}},\ }\href {\doibase
  10.3847/1538-4357/aaa1ed} {\bibfield  {journal} {\bibinfo  {journal}
  {Astrophys. J.}\ }\textbf {\bibinfo {volume} {853}},\ \bibinfo {pages} {119}
  (\bibinfo {year} {2018})},\ \Eprint {http://arxiv.org/abs/1707.06547}
  {arXiv:1707.06547 [astro-ph.CO]} \BibitemShut {NoStop}%
\bibitem [{\citenamefont {Wang}\ \emph {et~al.}(2017)\citenamefont {Wang},
  \citenamefont {Xu},\ and\ \citenamefont {Zhao}}]{Wang:2017yfu}%
  \BibitemOpen
  \bibfield  {author} {\bibinfo {author} {\bibfnamefont {Y.}~\bibnamefont
  {Wang}}, \bibinfo {author} {\bibfnamefont {L.}~\bibnamefont {Xu}}, \ and\
  \bibinfo {author} {\bibfnamefont {G.-B.}\ \bibnamefont {Zhao}},\ }\href
  {\doibase 10.3847/1538-4357/aa8f48} {\bibfield  {journal} {\bibinfo
  {journal} {Astrophys. J.}\ }\textbf {\bibinfo {volume} {849}},\ \bibinfo
  {pages} {84} (\bibinfo {year} {2017})},\ \Eprint
  {http://arxiv.org/abs/1706.09149} {arXiv:1706.09149 [astro-ph.CO]}
  \BibitemShut {NoStop}%
\bibitem [{\citenamefont {Cuceu}\ \emph {et~al.}(2019)\citenamefont {Cuceu},
  \citenamefont {Farr}, \citenamefont {Lemos},\ and\ \citenamefont
  {Font-Ribera}}]{Cuceu:2019for}%
  \BibitemOpen
  \bibfield  {author} {\bibinfo {author} {\bibfnamefont {A.}~\bibnamefont
  {Cuceu}}, \bibinfo {author} {\bibfnamefont {J.}~\bibnamefont {Farr}},
  \bibinfo {author} {\bibfnamefont {P.}~\bibnamefont {Lemos}}, \ and\ \bibinfo
  {author} {\bibfnamefont {A.}~\bibnamefont {Font-Ribera}},\ }\href {\doibase
  10.1088/1475-7516/2019/10/044} {\bibfield  {journal} {\bibinfo  {journal}
  {JCAP}\ }\textbf {\bibinfo {volume} {10}},\ \bibinfo {pages} {044} (\bibinfo
  {year} {2019})},\ \Eprint {http://arxiv.org/abs/1906.11628} {arXiv:1906.11628
  [astro-ph.CO]} \BibitemShut {NoStop}%
\bibitem [{\citenamefont {D'Amico}\ \emph
  {et~al.}(2020{\natexlab{a}})\citenamefont {D'Amico}, \citenamefont {Gleyzes},
  \citenamefont {Kokron}, \citenamefont {Markovic}, \citenamefont {Senatore},
  \citenamefont {Zhang}, \citenamefont {Beutler},\ and\ \citenamefont
  {Gil-Mar\'\i{}n}}]{DAmico:2019fhj}%
  \BibitemOpen
  \bibfield  {author} {\bibinfo {author} {\bibfnamefont {G.}~\bibnamefont
  {D'Amico}}, \bibinfo {author} {\bibfnamefont {J.}~\bibnamefont {Gleyzes}},
  \bibinfo {author} {\bibfnamefont {N.}~\bibnamefont {Kokron}}, \bibinfo
  {author} {\bibfnamefont {K.}~\bibnamefont {Markovic}}, \bibinfo {author}
  {\bibfnamefont {L.}~\bibnamefont {Senatore}}, \bibinfo {author}
  {\bibfnamefont {P.}~\bibnamefont {Zhang}}, \bibinfo {author} {\bibfnamefont
  {F.}~\bibnamefont {Beutler}}, \ and\ \bibinfo {author} {\bibfnamefont
  {H.}~\bibnamefont {Gil-Mar\'\i{}n}},\ }\href {\doibase
  10.1088/1475-7516/2020/05/005} {\bibfield  {journal} {\bibinfo  {journal}
  {JCAP}\ }\textbf {\bibinfo {volume} {05}},\ \bibinfo {pages} {005} (\bibinfo
  {year} {2020}{\natexlab{a}})},\ \Eprint {http://arxiv.org/abs/1909.05271}
  {arXiv:1909.05271 [astro-ph.CO]} \BibitemShut {NoStop}%
\bibitem [{\citenamefont {Ivanov}\ \emph
  {et~al.}(2020{\natexlab{a}})\citenamefont {Ivanov}, \citenamefont
  {Simonovi\'c},\ and\ \citenamefont {Zaldarriaga}}]{Ivanov:2019pdj}%
  \BibitemOpen
  \bibfield  {author} {\bibinfo {author} {\bibfnamefont {M.~M.}\ \bibnamefont
  {Ivanov}}, \bibinfo {author} {\bibfnamefont {M.}~\bibnamefont {Simonovi\'c}},
  \ and\ \bibinfo {author} {\bibfnamefont {M.}~\bibnamefont {Zaldarriaga}},\
  }\href {\doibase 10.1088/1475-7516/2020/05/042} {\bibfield  {journal}
  {\bibinfo  {journal} {JCAP}\ }\textbf {\bibinfo {volume} {05}},\ \bibinfo
  {pages} {042} (\bibinfo {year} {2020}{\natexlab{a}})},\ \Eprint
  {http://arxiv.org/abs/1909.05277} {arXiv:1909.05277 [astro-ph.CO]}
  \BibitemShut {NoStop}%
\bibitem [{\citenamefont {Philcox}\ \emph
  {et~al.}(2020{\natexlab{b}})\citenamefont {Philcox}, \citenamefont {Ivanov},
  \citenamefont {Simonovi\'c},\ and\ \citenamefont
  {Zaldarriaga}}]{Philcox:2020vvt}%
  \BibitemOpen
  \bibfield  {author} {\bibinfo {author} {\bibfnamefont {O.~H.}\ \bibnamefont
  {Philcox}}, \bibinfo {author} {\bibfnamefont {M.~M.}\ \bibnamefont {Ivanov}},
  \bibinfo {author} {\bibfnamefont {M.}~\bibnamefont {Simonovi\'c}}, \ and\
  \bibinfo {author} {\bibfnamefont {M.}~\bibnamefont {Zaldarriaga}},\ }\href
  {\doibase 10.1088/1475-7516/2020/05/032} {\bibfield  {journal} {\bibinfo
  {journal} {JCAP}\ }\textbf {\bibinfo {volume} {05}},\ \bibinfo {pages} {032}
  (\bibinfo {year} {2020}{\natexlab{b}})},\ \Eprint
  {http://arxiv.org/abs/2002.04035} {arXiv:2002.04035 [astro-ph.CO]}
  \BibitemShut {NoStop}%
\bibitem [{\citenamefont {Alam}\ \emph {et~al.}(2020)\citenamefont {Alam} \emph
  {et~al.}}]{Alam:2020sor}%
  \BibitemOpen
  \bibfield  {author} {\bibinfo {author} {\bibfnamefont {S.}~\bibnamefont
  {Alam}} \emph {et~al.} (\bibinfo {collaboration} {eBOSS}),\ }\href@noop {} {\
   (\bibinfo {year} {2020})},\ \Eprint {http://arxiv.org/abs/2007.08991}
  {arXiv:2007.08991 [astro-ph.CO]} \BibitemShut {NoStop}%
\bibitem [{\citenamefont {Cyburt}\ \emph {et~al.}(2016)\citenamefont {Cyburt},
  \citenamefont {Fields}, \citenamefont {Olive},\ and\ \citenamefont
  {Yeh}}]{Cyburt:2015mya}%
  \BibitemOpen
  \bibfield  {author} {\bibinfo {author} {\bibfnamefont {R.~H.}\ \bibnamefont
  {Cyburt}}, \bibinfo {author} {\bibfnamefont {B.~D.}\ \bibnamefont {Fields}},
  \bibinfo {author} {\bibfnamefont {K.~A.}\ \bibnamefont {Olive}}, \ and\
  \bibinfo {author} {\bibfnamefont {T.-H.}\ \bibnamefont {Yeh}},\ }\href
  {\doibase 10.1103/RevModPhys.88.015004} {\bibfield  {journal} {\bibinfo
  {journal} {Rev. Mod. Phys.}\ }\textbf {\bibinfo {volume} {88}},\ \bibinfo
  {pages} {015004} (\bibinfo {year} {2016})},\ \Eprint
  {http://arxiv.org/abs/1505.01076} {arXiv:1505.01076 [astro-ph.CO]}
  \BibitemShut {NoStop}%
\bibitem [{\citenamefont {Abbott}\ \emph
  {et~al.}(2018{\natexlab{a}})\citenamefont {Abbott} \emph
  {et~al.}}]{Abbott:2017smn}%
  \BibitemOpen
  \bibfield  {author} {\bibinfo {author} {\bibfnamefont {T.~M.~C.}\
  \bibnamefont {Abbott}} \emph {et~al.} (\bibinfo {collaboration} {DES}),\
  }\href {\doibase 10.1093/mnras/sty1939} {\bibfield  {journal} {\bibinfo
  {journal} {Mon. Not. Roy. Astron. Soc.}\ }\textbf {\bibinfo {volume} {480}},\
  \bibinfo {pages} {3879} (\bibinfo {year} {2018}{\natexlab{a}})},\ \Eprint
  {http://arxiv.org/abs/1711.00403} {arXiv:1711.00403 [astro-ph.CO]}
  \BibitemShut {NoStop}%
\bibitem [{\citenamefont {Aylor}\ \emph {et~al.}(2019)\citenamefont {Aylor},
  \citenamefont {Joy}, \citenamefont {Knox}, \citenamefont {Millea},
  \citenamefont {Raghunathan},\ and\ \citenamefont {Wu}}]{Aylor:2018drw}%
  \BibitemOpen
  \bibfield  {author} {\bibinfo {author} {\bibfnamefont {K.}~\bibnamefont
  {Aylor}}, \bibinfo {author} {\bibfnamefont {M.}~\bibnamefont {Joy}}, \bibinfo
  {author} {\bibfnamefont {L.}~\bibnamefont {Knox}}, \bibinfo {author}
  {\bibfnamefont {M.}~\bibnamefont {Millea}}, \bibinfo {author} {\bibfnamefont
  {S.}~\bibnamefont {Raghunathan}}, \ and\ \bibinfo {author} {\bibfnamefont
  {W.~K.}\ \bibnamefont {Wu}},\ }\href {\doibase 10.3847/1538-4357/ab0898}
  {\bibfield  {journal} {\bibinfo  {journal} {Astrophys. J.}\ }\textbf
  {\bibinfo {volume} {874}},\ \bibinfo {pages} {4} (\bibinfo {year} {2019})},\
  \Eprint {http://arxiv.org/abs/1811.00537} {arXiv:1811.00537 [astro-ph.CO]}
  \BibitemShut {NoStop}%
\bibitem [{\citenamefont {Wojtak}\ and\ \citenamefont
  {Agnello}(2019)}]{Wojtak:2019tkz}%
  \BibitemOpen
  \bibfield  {author} {\bibinfo {author} {\bibfnamefont {R.~a.}\ \bibnamefont
  {Wojtak}}\ and\ \bibinfo {author} {\bibfnamefont {A.}~\bibnamefont
  {Agnello}},\ }\href {\doibase 10.1093/mnras/stz1163} {\bibfield  {journal}
  {\bibinfo  {journal} {Mon. Not. Roy. Astron. Soc.}\ }\textbf {\bibinfo
  {volume} {486}},\ \bibinfo {pages} {5046} (\bibinfo {year} {2019})},\ \Eprint
  {http://arxiv.org/abs/1908.02401} {arXiv:1908.02401 [astro-ph.CO]}
  \BibitemShut {NoStop}%
\bibitem [{\citenamefont {Arendse}\ \emph {et~al.}(2019)\citenamefont {Arendse}
  \emph {et~al.}}]{Arendse:2019hev}%
  \BibitemOpen
  \bibfield  {author} {\bibinfo {author} {\bibfnamefont {N.}~\bibnamefont
  {Arendse}} \emph {et~al.},\ }\href@noop {} {\  (\bibinfo {year} {2019})},\
  \Eprint {http://arxiv.org/abs/1909.07986} {arXiv:1909.07986 [astro-ph.CO]}
  \BibitemShut {NoStop}%
\bibitem [{\citenamefont {Baxter}\ and\ \citenamefont
  {Sherwin}(2020)}]{Baxter:2020qlr}%
  \BibitemOpen
  \bibfield  {author} {\bibinfo {author} {\bibfnamefont {E.~J.}\ \bibnamefont
  {Baxter}}\ and\ \bibinfo {author} {\bibfnamefont {B.~D.}\ \bibnamefont
  {Sherwin}},\ }\href@noop {} {\  (\bibinfo {year} {2020})},\ \Eprint
  {http://arxiv.org/abs/2007.04007} {arXiv:2007.04007 [astro-ph.CO]}
  \BibitemShut {NoStop}%
\bibitem [{\citenamefont {Zhang}\ and\ \citenamefont
  {Huang}(2020)}]{Zhang:2020uan}%
  \BibitemOpen
  \bibfield  {author} {\bibinfo {author} {\bibfnamefont {X.}~\bibnamefont
  {Zhang}}\ and\ \bibinfo {author} {\bibfnamefont {Q.-G.}\ \bibnamefont
  {Huang}},\ }\href@noop {} {\  (\bibinfo {year} {2020})},\ \Eprint
  {http://arxiv.org/abs/2006.16692} {arXiv:2006.16692 [astro-ph.CO]}
  \BibitemShut {NoStop}%
\bibitem [{\citenamefont {Magana}\ \emph {et~al.}(2018)\citenamefont {Magana},
  \citenamefont {Amante}, \citenamefont {Garcia-Aspeitia},\ and\ \citenamefont
  {Motta}}]{Magana:2017nfs}%
  \BibitemOpen
  \bibfield  {author} {\bibinfo {author} {\bibfnamefont {J.}~\bibnamefont
  {Magana}}, \bibinfo {author} {\bibfnamefont {M.~H.}\ \bibnamefont {Amante}},
  \bibinfo {author} {\bibfnamefont {M.~A.}\ \bibnamefont {Garcia-Aspeitia}}, \
  and\ \bibinfo {author} {\bibfnamefont {V.}~\bibnamefont {Motta}},\ }\href
  {\doibase 10.1093/mnras/sty260} {\bibfield  {journal} {\bibinfo  {journal}
  {Mon. Not. Roy. Astron. Soc.}\ }\textbf {\bibinfo {volume} {476}},\ \bibinfo
  {pages} {1036} (\bibinfo {year} {2018})},\ \Eprint
  {http://arxiv.org/abs/1706.09848} {arXiv:1706.09848 [astro-ph.CO]}
  \BibitemShut {NoStop}%
\bibitem [{\citenamefont {Ade}\ \emph {et~al.}(2019)\citenamefont {Ade} \emph
  {et~al.}}]{Ade:2018sbj}%
  \BibitemOpen
  \bibfield  {author} {\bibinfo {author} {\bibfnamefont {P.}~\bibnamefont
  {Ade}} \emph {et~al.} (\bibinfo {collaboration} {Simons Observatory}),\
  }\href {\doibase 10.1088/1475-7516/2019/02/056} {\bibfield  {journal}
  {\bibinfo  {journal} {JCAP}\ }\textbf {\bibinfo {volume} {02}},\ \bibinfo
  {pages} {056} (\bibinfo {year} {2019})},\ \Eprint
  {http://arxiv.org/abs/1808.07445} {arXiv:1808.07445 [astro-ph.CO]}
  \BibitemShut {NoStop}%
\bibitem [{\citenamefont {Abazajian}\ \emph {et~al.}(2016)\citenamefont
  {Abazajian} \emph {et~al.}}]{Abazajian:2016yjj}%
  \BibitemOpen
  \bibfield  {author} {\bibinfo {author} {\bibfnamefont {K.~N.}\ \bibnamefont
  {Abazajian}} \emph {et~al.} (\bibinfo {collaboration} {CMB-S4}),\ }\href@noop
  {} {\  (\bibinfo {year} {2016})},\ \Eprint {http://arxiv.org/abs/1610.02743}
  {arXiv:1610.02743 [astro-ph.CO]} \BibitemShut {NoStop}%
\bibitem [{\citenamefont {Mirmelstein}\ \emph {et~al.}(2019)\citenamefont
  {Mirmelstein}, \citenamefont {Carron},\ and\ \citenamefont
  {Lewis}}]{Mirmelstein:2019sxi}%
  \BibitemOpen
  \bibfield  {author} {\bibinfo {author} {\bibfnamefont {M.}~\bibnamefont
  {Mirmelstein}}, \bibinfo {author} {\bibfnamefont {J.}~\bibnamefont {Carron}},
  \ and\ \bibinfo {author} {\bibfnamefont {A.}~\bibnamefont {Lewis}},\ }\href
  {\doibase 10.1103/PhysRevD.100.123509} {\bibfield  {journal} {\bibinfo
  {journal} {Phys. Rev. D}\ }\textbf {\bibinfo {volume} {100}},\ \bibinfo
  {pages} {123509} (\bibinfo {year} {2019})},\ \Eprint
  {http://arxiv.org/abs/1909.02653} {arXiv:1909.02653 [astro-ph.CO]}
  \BibitemShut {NoStop}%
\bibitem [{euc()}]{euclid}%
  \BibitemOpen
  \href@noop {} {}\bibinfo {note} {\url{http://www.euclid-ec.org}}\BibitemShut
  {NoStop}%
\bibitem [{lss()}]{lsst}%
  \BibitemOpen
  \href@noop {} {}\bibinfo {note} {\url{http://www.lsst.org}}\BibitemShut
  {NoStop}%
\bibitem [{\citenamefont {Anderson}\ \emph {et~al.}(2014)\citenamefont
  {Anderson} \emph {et~al.}}]{Anderson:2013zyy}%
  \BibitemOpen
  \bibfield  {author} {\bibinfo {author} {\bibfnamefont {L.}~\bibnamefont
  {Anderson}} \emph {et~al.} (\bibinfo {collaboration} {BOSS}),\ }\href
  {\doibase 10.1093/mnras/stu523} {\bibfield  {journal} {\bibinfo  {journal}
  {Mon. Not. Roy. Astron. Soc.}\ }\textbf {\bibinfo {volume} {441}},\ \bibinfo
  {pages} {24} (\bibinfo {year} {2014})},\ \Eprint
  {http://arxiv.org/abs/1312.4877} {arXiv:1312.4877 [astro-ph.CO]} \BibitemShut
  {NoStop}%
\bibitem [{\citenamefont {Aubourg}\ \emph {et~al.}(2015)\citenamefont {Aubourg}
  \emph {et~al.}}]{Aubourg:2014yra}%
  \BibitemOpen
  \bibfield  {author} {\bibinfo {author} {\bibfnamefont {E.}~\bibnamefont
  {Aubourg}} \emph {et~al.},\ }\href {\doibase 10.1103/PhysRevD.92.123516}
  {\bibfield  {journal} {\bibinfo  {journal} {Phys. Rev. D}\ }\textbf {\bibinfo
  {volume} {92}},\ \bibinfo {pages} {123516} (\bibinfo {year} {2015})},\
  \Eprint {http://arxiv.org/abs/1411.1074} {arXiv:1411.1074 [astro-ph.CO]}
  \BibitemShut {NoStop}%
\bibitem [{\citenamefont {Eisenstein}\ \emph {et~al.}(2005)\citenamefont
  {Eisenstein} \emph {et~al.}}]{Eisenstein:2005su}%
  \BibitemOpen
  \bibfield  {author} {\bibinfo {author} {\bibfnamefont {D.~J.}\ \bibnamefont
  {Eisenstein}} \emph {et~al.} (\bibinfo {collaboration} {SDSS}),\ }\href
  {\doibase 10.1086/466512} {\bibfield  {journal} {\bibinfo  {journal}
  {Astrophys. J.}\ }\textbf {\bibinfo {volume} {633}},\ \bibinfo {pages} {560}
  (\bibinfo {year} {2005})},\ \Eprint {http://arxiv.org/abs/astro-ph/0501171}
  {arXiv:astro-ph/0501171} \BibitemShut {NoStop}%
\bibitem [{\citenamefont {Zhao}\ \emph {et~al.}(2020)\citenamefont {Zhao} \emph
  {et~al.}}]{Zhao:2020tis}%
  \BibitemOpen
  \bibfield  {author} {\bibinfo {author} {\bibfnamefont {G.-B.}\ \bibnamefont
  {Zhao}} \emph {et~al.},\ }\href@noop {} {\  (\bibinfo {year} {2020})},\
  \Eprint {http://arxiv.org/abs/2007.09011} {arXiv:2007.09011 [astro-ph.CO]}
  \BibitemShut {NoStop}%
\bibitem [{\citenamefont {Wang}\ \emph {et~al.}(2020)\citenamefont {Wang} \emph
  {et~al.}}]{Wang:2020tje}%
  \BibitemOpen
  \bibfield  {author} {\bibinfo {author} {\bibfnamefont {Y.}~\bibnamefont
  {Wang}} \emph {et~al.},\ }\href {\doibase 10.1093/mnras/staa2593} {\
  (\bibinfo {year} {2020}),\ 10.1093/mnras/staa2593},\ \Eprint
  {http://arxiv.org/abs/2007.09010} {arXiv:2007.09010 [astro-ph.CO]}
  \BibitemShut {NoStop}%
\bibitem [{\citenamefont {Hou}\ \emph {et~al.}(2020)\citenamefont {Hou} \emph
  {et~al.}}]{Hou:2020rse}%
  \BibitemOpen
  \bibfield  {author} {\bibinfo {author} {\bibfnamefont {J.}~\bibnamefont
  {Hou}} \emph {et~al.},\ }\href@noop {} {\  (\bibinfo {year} {2020})},\
  \Eprint {http://arxiv.org/abs/2007.08998} {arXiv:2007.08998 [astro-ph.CO]}
  \BibitemShut {NoStop}%
\bibitem [{\citenamefont {du~Mas~des Bourboux}\ \emph
  {et~al.}(2020)\citenamefont {du~Mas~des Bourboux} \emph
  {et~al.}}]{duMasdesBourboux:2020pck}%
  \BibitemOpen
  \bibfield  {author} {\bibinfo {author} {\bibfnamefont {H.}~\bibnamefont
  {du~Mas~des Bourboux}} \emph {et~al.},\ }\href@noop {} {\  (\bibinfo {year}
  {2020})},\ \Eprint {http://arxiv.org/abs/2007.08995} {arXiv:2007.08995
  [astro-ph.CO]} \BibitemShut {NoStop}%
\bibitem [{\citenamefont {Beutler}\ \emph {et~al.}(2011)\citenamefont
  {Beutler}, \citenamefont {Blake}, \citenamefont {Colless}, \citenamefont
  {Jones}, \citenamefont {Staveley-Smith}, \citenamefont {Campbell},
  \citenamefont {Parker}, \citenamefont {Saunders},\ and\ \citenamefont
  {Watson}}]{Beutler:2011hx}%
  \BibitemOpen
  \bibfield  {author} {\bibinfo {author} {\bibfnamefont {F.}~\bibnamefont
  {Beutler}}, \bibinfo {author} {\bibfnamefont {C.}~\bibnamefont {Blake}},
  \bibinfo {author} {\bibfnamefont {M.}~\bibnamefont {Colless}}, \bibinfo
  {author} {\bibfnamefont {D.}~\bibnamefont {Jones}}, \bibinfo {author}
  {\bibfnamefont {L.}~\bibnamefont {Staveley-Smith}}, \bibinfo {author}
  {\bibfnamefont {L.}~\bibnamefont {Campbell}}, \bibinfo {author}
  {\bibfnamefont {Q.}~\bibnamefont {Parker}}, \bibinfo {author} {\bibfnamefont
  {W.}~\bibnamefont {Saunders}}, \ and\ \bibinfo {author} {\bibfnamefont
  {F.}~\bibnamefont {Watson}},\ }\href {\doibase
  10.1111/j.1365-2966.2011.19250.x} {\bibfield  {journal} {\bibinfo  {journal}
  {Mon. Not. Roy. Astron. Soc.}\ }\textbf {\bibinfo {volume} {416}},\ \bibinfo
  {pages} {3017} (\bibinfo {year} {2011})},\ \Eprint
  {http://arxiv.org/abs/1106.3366} {arXiv:1106.3366 [astro-ph.CO]} \BibitemShut
  {NoStop}%
\bibitem [{\citenamefont {Ross}\ \emph {et~al.}(2015)\citenamefont {Ross},
  \citenamefont {Samushia}, \citenamefont {Howlett}, \citenamefont {Percival},
  \citenamefont {Burden},\ and\ \citenamefont {Manera}}]{Ross:2014qpa}%
  \BibitemOpen
  \bibfield  {author} {\bibinfo {author} {\bibfnamefont {A.~J.}\ \bibnamefont
  {Ross}}, \bibinfo {author} {\bibfnamefont {L.}~\bibnamefont {Samushia}},
  \bibinfo {author} {\bibfnamefont {C.}~\bibnamefont {Howlett}}, \bibinfo
  {author} {\bibfnamefont {W.~J.}\ \bibnamefont {Percival}}, \bibinfo {author}
  {\bibfnamefont {A.}~\bibnamefont {Burden}}, \ and\ \bibinfo {author}
  {\bibfnamefont {M.}~\bibnamefont {Manera}},\ }\href {\doibase
  10.1093/mnras/stv154} {\bibfield  {journal} {\bibinfo  {journal} {Mon. Not.
  Roy. Astron. Soc.}\ }\textbf {\bibinfo {volume} {449}},\ \bibinfo {pages}
  {835} (\bibinfo {year} {2015})},\ \Eprint {http://arxiv.org/abs/1409.3242}
  {arXiv:1409.3242 [astro-ph.CO]} \BibitemShut {NoStop}%
\bibitem [{\citenamefont {Abbott}\ \emph
  {et~al.}(2018{\natexlab{b}})\citenamefont {Abbott} \emph
  {et~al.}}]{Abbott:2017wau}%
  \BibitemOpen
  \bibfield  {author} {\bibinfo {author} {\bibfnamefont {T.~M.~C.}\
  \bibnamefont {Abbott}} \emph {et~al.} (\bibinfo {collaboration} {DES}),\
  }\href {\doibase 10.1103/PhysRevD.98.043526} {\bibfield  {journal} {\bibinfo
  {journal} {Phys. Rev.}\ }\textbf {\bibinfo {volume} {D98}},\ \bibinfo {pages}
  {043526} (\bibinfo {year} {2018}{\natexlab{b}})},\ \Eprint
  {http://arxiv.org/abs/1708.01530} {arXiv:1708.01530 [astro-ph.CO]}
  \BibitemShut {NoStop}%
\bibitem [{\citenamefont {Aghanim}\ \emph
  {et~al.}(2018{\natexlab{b}})\citenamefont {Aghanim} \emph
  {et~al.}}]{Aghanim:2018oex}%
  \BibitemOpen
  \bibfield  {author} {\bibinfo {author} {\bibfnamefont {N.}~\bibnamefont
  {Aghanim}} \emph {et~al.} (\bibinfo {collaboration} {Planck}),\ }\href@noop
  {} {\  (\bibinfo {year} {2018}{\natexlab{b}})},\ \Eprint
  {http://arxiv.org/abs/1807.06210} {arXiv:1807.06210 [astro-ph.CO]}
  \BibitemShut {NoStop}%
\bibitem [{\citenamefont {Wu}\ \emph {et~al.}(2019)\citenamefont {Wu} \emph
  {et~al.}}]{Wu:2019hek}%
  \BibitemOpen
  \bibfield  {author} {\bibinfo {author} {\bibfnamefont {W.}~\bibnamefont {Wu}}
  \emph {et~al.},\ }\href {\doibase 10.3847/1538-4357/ab4186} {\bibfield
  {journal} {\bibinfo  {journal} {Astrophys. J.}\ }\textbf {\bibinfo {volume}
  {884}},\ \bibinfo {pages} {70} (\bibinfo {year} {2019})},\ \Eprint
  {http://arxiv.org/abs/1905.05777} {arXiv:1905.05777 [astro-ph.CO]}
  \BibitemShut {NoStop}%
\bibitem [{\citenamefont {Bianchini}\ \emph {et~al.}(2020)\citenamefont
  {Bianchini} \emph {et~al.}}]{Bianchini:2019vxp}%
  \BibitemOpen
  \bibfield  {author} {\bibinfo {author} {\bibfnamefont {F.}~\bibnamefont
  {Bianchini}} \emph {et~al.} (\bibinfo {collaboration} {SPT}),\ }\href
  {\doibase 10.3847/1538-4357/ab6082} {\bibfield  {journal} {\bibinfo
  {journal} {Astrophys. J.}\ }\textbf {\bibinfo {volume} {888}},\ \bibinfo
  {pages} {119} (\bibinfo {year} {2020})},\ \Eprint
  {http://arxiv.org/abs/1910.07157} {arXiv:1910.07157 [astro-ph.CO]}
  \BibitemShut {NoStop}%
\bibitem [{\citenamefont {{Moresco}}\ \emph {et~al.}(2016)\citenamefont
  {{Moresco}}, \citenamefont {{Pozzetti}}, \citenamefont {{Cimatti}},
  \citenamefont {{Jimenez}}, \citenamefont {{Maraston}}, \citenamefont
  {{Verde}}, \citenamefont {{Thomas}}, \citenamefont {{Citro}}, \citenamefont
  {{Tojeiro}},\ and\ \citenamefont {{Wilkinson}}}]{OHD:Moresco_2016}%
  \BibitemOpen
  \bibfield  {author} {\bibinfo {author} {\bibfnamefont {M.}~\bibnamefont
  {{Moresco}}}, \bibinfo {author} {\bibfnamefont {L.}~\bibnamefont
  {{Pozzetti}}}, \bibinfo {author} {\bibfnamefont {A.}~\bibnamefont
  {{Cimatti}}}, \bibinfo {author} {\bibfnamefont {R.}~\bibnamefont
  {{Jimenez}}}, \bibinfo {author} {\bibfnamefont {C.}~\bibnamefont
  {{Maraston}}}, \bibinfo {author} {\bibfnamefont {L.}~\bibnamefont {{Verde}}},
  \bibinfo {author} {\bibfnamefont {D.}~\bibnamefont {{Thomas}}}, \bibinfo
  {author} {\bibfnamefont {A.}~\bibnamefont {{Citro}}}, \bibinfo {author}
  {\bibfnamefont {R.}~\bibnamefont {{Tojeiro}}}, \ and\ \bibinfo {author}
  {\bibfnamefont {D.}~\bibnamefont {{Wilkinson}}},\ }\href {\doibase
  10.1088/1475-7516/2016/05/014} {\bibfield  {journal} {\bibinfo  {journal}
  {JCAP}\ }\textbf {\bibinfo {volume} {2016}},\ \bibinfo {eid} {014} (\bibinfo
  {year} {2016})},\ \Eprint {http://arxiv.org/abs/1601.01701} {arXiv:1601.01701
  [astro-ph.CO]} \BibitemShut {NoStop}%
\bibitem [{\citenamefont {Ratsimbazafy}\ \emph {et~al.}(2017)\citenamefont
  {Ratsimbazafy}, \citenamefont {Loubser}, \citenamefont {Crawford},
  \citenamefont {Cress}, \citenamefont {Bassett}, \citenamefont {Nichol},\ and\
  \citenamefont {Vaisanen}}]{Ratsimbazafy:2017vga}%
  \BibitemOpen
  \bibfield  {author} {\bibinfo {author} {\bibfnamefont {A.}~\bibnamefont
  {Ratsimbazafy}}, \bibinfo {author} {\bibfnamefont {S.}~\bibnamefont
  {Loubser}}, \bibinfo {author} {\bibfnamefont {S.}~\bibnamefont {Crawford}},
  \bibinfo {author} {\bibfnamefont {C.}~\bibnamefont {Cress}}, \bibinfo
  {author} {\bibfnamefont {B.}~\bibnamefont {Bassett}}, \bibinfo {author}
  {\bibfnamefont {R.}~\bibnamefont {Nichol}}, \ and\ \bibinfo {author}
  {\bibfnamefont {P.}~\bibnamefont {Vaisanen}},\ }\href {\doibase
  10.1093/mnras/stx301} {\bibfield  {journal} {\bibinfo  {journal} {Mon. Not.
  Roy. Astron. Soc.}\ }\textbf {\bibinfo {volume} {467}},\ \bibinfo {pages}
  {3239} (\bibinfo {year} {2017})},\ \Eprint {http://arxiv.org/abs/1702.00418}
  {arXiv:1702.00418 [astro-ph.CO]} \BibitemShut {NoStop}%
\bibitem [{\citenamefont {Lewis}\ and\ \citenamefont
  {Bridle}(2002)}]{Lewis:2002ah}%
  \BibitemOpen
  \bibfield  {author} {\bibinfo {author} {\bibfnamefont {A.}~\bibnamefont
  {Lewis}}\ and\ \bibinfo {author} {\bibfnamefont {S.}~\bibnamefont {Bridle}},\
  }\href {\doibase 10.1103/PhysRevD.66.103511} {\bibfield  {journal} {\bibinfo
  {journal} {\prd}\ }\textbf {\bibinfo {volume} {66}},\ \bibinfo {pages}
  {103511} (\bibinfo {year} {2002})},\ \Eprint
  {http://arxiv.org/abs/astro-ph/0205436} {arXiv:astro-ph/0205436 [astro-ph]}
  \BibitemShut {NoStop}%
\bibitem [{\citenamefont {Ade}\ \emph {et~al.}(2016)\citenamefont {Ade} \emph
  {et~al.}}]{Ade:2015zua}%
  \BibitemOpen
  \bibfield  {author} {\bibinfo {author} {\bibfnamefont {P.}~\bibnamefont
  {Ade}} \emph {et~al.} (\bibinfo {collaboration} {Planck}),\ }\href {\doibase
  10.1051/0004-6361/201525941} {\bibfield  {journal} {\bibinfo  {journal}
  {Astron. Astrophys.}\ }\textbf {\bibinfo {volume} {594}},\ \bibinfo {pages}
  {A15} (\bibinfo {year} {2016})},\ \Eprint {http://arxiv.org/abs/1502.01591}
  {arXiv:1502.01591 [astro-ph.CO]} \BibitemShut {NoStop}%
\bibitem [{\citenamefont {Scolnic}\ \emph {et~al.}(2018)\citenamefont {Scolnic}
  \emph {et~al.}}]{Scolnic:2017caz}%
  \BibitemOpen
  \bibfield  {author} {\bibinfo {author} {\bibfnamefont {D.}~\bibnamefont
  {Scolnic}} \emph {et~al.},\ }\href {\doibase 10.3847/1538-4357/aab9bb}
  {\bibfield  {journal} {\bibinfo  {journal} {Astrophys. J.}\ }\textbf
  {\bibinfo {volume} {859}},\ \bibinfo {pages} {101} (\bibinfo {year}
  {2018})},\ \Eprint {http://arxiv.org/abs/1710.00845} {arXiv:1710.00845
  [astro-ph.CO]} \BibitemShut {NoStop}%
\bibitem [{\citenamefont {Percival}\ \emph {et~al.}(2002)\citenamefont
  {Percival} \emph {et~al.}}]{Percival:2002gq}%
  \BibitemOpen
  \bibfield  {author} {\bibinfo {author} {\bibfnamefont {W.~J.}\ \bibnamefont
  {Percival}} \emph {et~al.} (\bibinfo {collaboration} {2dFGRS Team}),\ }\href
  {\doibase 10.1046/j.1365-8711.2002.06001.x} {\bibfield  {journal} {\bibinfo
  {journal} {Mon. Not. Roy. Astron. Soc.}\ }\textbf {\bibinfo {volume} {337}},\
  \bibinfo {pages} {1068} (\bibinfo {year} {2002})},\ \Eprint
  {http://arxiv.org/abs/astro-ph/0206256} {arXiv:astro-ph/0206256} \BibitemShut
  {NoStop}%
\bibitem [{\citenamefont {Hill}\ \emph {et~al.}(2020)\citenamefont {Hill},
  \citenamefont {McDonough}, \citenamefont {Toomey},\ and\ \citenamefont
  {Alexander}}]{Hill:2020osr}%
  \BibitemOpen
  \bibfield  {author} {\bibinfo {author} {\bibfnamefont {J.~C.}\ \bibnamefont
  {Hill}}, \bibinfo {author} {\bibfnamefont {E.}~\bibnamefont {McDonough}},
  \bibinfo {author} {\bibfnamefont {M.~W.}\ \bibnamefont {Toomey}}, \ and\
  \bibinfo {author} {\bibfnamefont {S.}~\bibnamefont {Alexander}},\ }\href
  {\doibase 10.1103/PhysRevD.102.043507} {\bibfield  {journal} {\bibinfo
  {journal} {Phys. Rev. D}\ }\textbf {\bibinfo {volume} {102}},\ \bibinfo
  {pages} {043507} (\bibinfo {year} {2020})},\ \Eprint
  {http://arxiv.org/abs/2003.07355} {arXiv:2003.07355 [astro-ph.CO]}
  \BibitemShut {NoStop}%
\bibitem [{\citenamefont {Ivanov}\ \emph
  {et~al.}(2020{\natexlab{b}})\citenamefont {Ivanov}, \citenamefont
  {McDonough}, \citenamefont {Hill}, \citenamefont {Simonovi\'c}, \citenamefont
  {Toomey}, \citenamefont {Alexander},\ and\ \citenamefont
  {Zaldarriaga}}]{Ivanov:2020ril}%
  \BibitemOpen
  \bibfield  {author} {\bibinfo {author} {\bibfnamefont {M.~M.}\ \bibnamefont
  {Ivanov}}, \bibinfo {author} {\bibfnamefont {E.}~\bibnamefont {McDonough}},
  \bibinfo {author} {\bibfnamefont {J.~C.}\ \bibnamefont {Hill}}, \bibinfo
  {author} {\bibfnamefont {M.}~\bibnamefont {Simonovi\'c}}, \bibinfo {author}
  {\bibfnamefont {M.~W.}\ \bibnamefont {Toomey}}, \bibinfo {author}
  {\bibfnamefont {S.}~\bibnamefont {Alexander}}, \ and\ \bibinfo {author}
  {\bibfnamefont {M.}~\bibnamefont {Zaldarriaga}},\ }\href@noop {} {\
  (\bibinfo {year} {2020}{\natexlab{b}})},\ \Eprint
  {http://arxiv.org/abs/2006.11235} {arXiv:2006.11235 [astro-ph.CO]}
  \BibitemShut {NoStop}%
\bibitem [{\citenamefont {D'Amico}\ \emph
  {et~al.}(2020{\natexlab{b}})\citenamefont {D'Amico}, \citenamefont
  {Senatore}, \citenamefont {Zhang},\ and\ \citenamefont
  {Zheng}}]{DAmico:2020ods}%
  \BibitemOpen
  \bibfield  {author} {\bibinfo {author} {\bibfnamefont {G.}~\bibnamefont
  {D'Amico}}, \bibinfo {author} {\bibfnamefont {L.}~\bibnamefont {Senatore}},
  \bibinfo {author} {\bibfnamefont {P.}~\bibnamefont {Zhang}}, \ and\ \bibinfo
  {author} {\bibfnamefont {H.}~\bibnamefont {Zheng}},\ }\href@noop {} {\
  (\bibinfo {year} {2020}{\natexlab{b}})},\ \Eprint
  {http://arxiv.org/abs/2006.12420} {arXiv:2006.12420 [astro-ph.CO]}
  \BibitemShut {NoStop}%
\bibitem [{\citenamefont {Ye}\ and\ \citenamefont {Piao}(2020)}]{Ye:2020oix}%
  \BibitemOpen
  \bibfield  {author} {\bibinfo {author} {\bibfnamefont {G.}~\bibnamefont
  {Ye}}\ and\ \bibinfo {author} {\bibfnamefont {Y.-S.}\ \bibnamefont {Piao}},\
  }\href@noop {} {\  (\bibinfo {year} {2020})},\ \Eprint
  {http://arxiv.org/abs/2008.10832} {arXiv:2008.10832 [astro-ph.CO]}
  \BibitemShut {NoStop}%
\bibitem [{\citenamefont {Murgia}\ \emph {et~al.}(2020)\citenamefont {Murgia},
  \citenamefont {Abell\'an},\ and\ \citenamefont {Poulin}}]{Murgia:2020ryi}%
  \BibitemOpen
  \bibfield  {author} {\bibinfo {author} {\bibfnamefont {R.}~\bibnamefont
  {Murgia}}, \bibinfo {author} {\bibfnamefont {G.~F.}\ \bibnamefont
  {Abell\'an}}, \ and\ \bibinfo {author} {\bibfnamefont {V.}~\bibnamefont
  {Poulin}},\ }\href@noop {} {\  (\bibinfo {year} {2020})},\ \Eprint
  {http://arxiv.org/abs/2009.10733} {arXiv:2009.10733 [astro-ph.CO]}
  \BibitemShut {NoStop}%
\bibitem [{\citenamefont {Smith}\ \emph {et~al.}(2020)\citenamefont {Smith},
  \citenamefont {Poulin}, \citenamefont {Bernal}, \citenamefont {Boddy},
  \citenamefont {Kamionkowski},\ and\ \citenamefont {Murgia}}]{Smith:2020rxx}%
  \BibitemOpen
  \bibfield  {author} {\bibinfo {author} {\bibfnamefont {T.~L.}\ \bibnamefont
  {Smith}}, \bibinfo {author} {\bibfnamefont {V.}~\bibnamefont {Poulin}},
  \bibinfo {author} {\bibfnamefont {J.~L.}\ \bibnamefont {Bernal}}, \bibinfo
  {author} {\bibfnamefont {K.~K.}\ \bibnamefont {Boddy}}, \bibinfo {author}
  {\bibfnamefont {M.}~\bibnamefont {Kamionkowski}}, \ and\ \bibinfo {author}
  {\bibfnamefont {R.}~\bibnamefont {Murgia}},\ }\href@noop {} {\  (\bibinfo
  {year} {2020})},\ \Eprint {http://arxiv.org/abs/2009.10740} {arXiv:2009.10740
  [astro-ph.CO]} \BibitemShut {NoStop}%
\bibitem [{\citenamefont {Aghamousa}\ \emph {et~al.}(2016)\citenamefont
  {Aghamousa} \emph {et~al.}}]{DESI}%
  \BibitemOpen
  \bibfield  {author} {\bibinfo {author} {\bibfnamefont {A.}~\bibnamefont
  {Aghamousa}} \emph {et~al.} (\bibinfo {collaboration} {DESI}),\ }\href@noop
  {} {\  (\bibinfo {year} {2016})},\ \Eprint {http://arxiv.org/abs/1611.00036}
  {arXiv:1611.00036 [astro-ph.IM]} \BibitemShut {NoStop}%
\bibitem [{\citenamefont {Freedman}\ \emph {et~al.}(2019)\citenamefont
  {Freedman} \emph {et~al.}}]{Freedman:2019jwv}%
  \BibitemOpen
  \bibfield  {author} {\bibinfo {author} {\bibfnamefont {W.~L.}\ \bibnamefont
  {Freedman}} \emph {et~al.},\ }\href {\doibase 10.3847/1538-4357/ab2f73} {\
  (\bibinfo {year} {2019}),\ 10.3847/1538-4357/ab2f73},\ \Eprint
  {http://arxiv.org/abs/1907.05922} {arXiv:1907.05922 [astro-ph.CO]}
  \BibitemShut {NoStop}%
\bibitem [{\citenamefont {Lewis}(2019)}]{Lewis:2019xzd}%
  \BibitemOpen
  \bibfield  {author} {\bibinfo {author} {\bibfnamefont {A.}~\bibnamefont
  {Lewis}},\ }\href {https://getdist.readthedocs.io} {\  (\bibinfo {year}
  {2019})},\ \Eprint {http://arxiv.org/abs/1910.13970} {arXiv:1910.13970
  [astro-ph.IM]} \BibitemShut {NoStop}%
\end{thebibliography}
%

\end{document}